\documentclass[12pt,letterpaper,prd,nofootinbib,floatfix,preprintnumbers]{revtex4}
\usepackage{graphicx}
\usepackage{amssymb}

\newcommand{\etmiss}{\not\hskip-5truedd E_{T} }
\begin{document}

\preprint{MSUHEP-060915}
\preprint{hep-ph/0609179}

\title{Phenomenology of Littlest Higgs Model with T-parity:\\
     including effects of T-odd fermions}

\author{Alexander Belyaev, Chuan-Ren Chen, Kazuhiro Tobe, C.-P. Yuan}
\affiliation{Department of Physics and Astronomy,
Michigan State University,
East Lansing, MI 48824, USA}

\begin{abstract}
We study the collider phenomenology of a Littlest Higgs model with 
T-parity. We first stress the important role of the T-odd $SU(2)$ 
doublet fermions (introduced to make the model T-parity invariant) 
in high energy scattering processes, such as 
$q\bar{q}\rightarrow W_H^+ W_H^-$ where $W_H^\pm$ are the T-odd 
partners of $W$-bosons. Because the mass of the T-odd $SU(2)$ 
doublet fermions cannot be too heavy to be consistent with low energy data,
they can be copiously produced 
at the CERN Large Hadron Collider (LHC). Therefore, we study the 
collider phenomenology of the model with emphasis on the contributions 
of the T-odd fermion to the production of the heavy T-parity partners
(either bosons or fermions) of the usual particles at the LHC. 
The production cross sections  and the decay branching ratios of the 
new heavy particles are classified and various experimental signatures 
are discussed. 
\end{abstract}

\date{\today}

\maketitle

\section{Introduction}
The standard model (SM) is an excellent low energy description of the elementary
particles. The absence of any significant deviations from the SM  predictions on
the electroweak precision measurements suggests that the cutoff scale of the SM,
as a low energy effective theory, is as large as, or larger than, 
$10$~TeV~\cite{Barbieri:2000gf}. However,  having such a relatively high cutoff scale
in the SM, the Higgs boson receives a large radiative correction  to its mass
parameter and therefore, the SM requires unsatisfactory fine-tuning to yield a
correct scale of the electroweak symmetry breaking. This fine-tuning problem of
the Higgs mass parameter (known as the ``Little hierarchy problem'') has been one
of the driving forces to consider physics beyond the SM.  Moreover, a recent
finding of the necessity of dark  matter candidate also provides a strong
motivation to seek for physics beyond the SM.

It has been shown  recently that the collective symmetry breaking mechanism in the
Little Higgs models~\cite{LittleHiggs} can provide an interesting solution to the
Little hierarchy problem and  the Littlest Higgs (LH) ~\cite{ Arkani-Hamed:2002qy}
model is the most economical  Little Higgs model discussed in the literature.  
However, the original version of the LH model suffers from precision 
electroweak constrains \cite{EWC} and the value of $f$, which characterizes
the mass scale of new particles in the model, is forced to  be larger than
about $4$ TeV.
Since the
cutoff scale of the model is about $4 \pi f$, the fine tuning
between the cutoff scale and the weak scale will be needed again
for a too large value of $f$.  The Littlest
Higgs model with  T-parity
(LHT)~\cite{Low:2004xc,Hubisz:2004ft,Hubisz:2005tx,Cheng:2003ju} is  one of the
attractive Little Higgs models. It provides a possible dark matter  candidate
\cite{Asano:2006nr} and  furthermore, all dangerous tree-level contributions to low
energy electroweak (EW) observables are  forbidden by T-parity and hence the
corrections to low energy EW observables  are loop-suppressed and small. As a
result, the relatively low new particle mass scale $f$ is still allowed by data,
e.g., $f>500$ GeV~\cite{Hubisz:2005tx}.

The LHT predicts heavy T-odd gauge bosons which are T-parity 
partners of the SM gauge bosons. Moreover, in order to implement  T-parity in the fermion sector, 
one  introduces the heavy
T-odd $SU(2)$-doublet fermions, which are T-parity partners of the SM
$SU(2)$-doublet fermions and
unique to Little Higgs models with T-parity.
Therefore, having the relatively low new particle mass scale $f$, the CERN Large Hadron Collider (LHC)
will have a great potential to directly produce the T-parity partners of the SM particles,
and hence it is important to probe the LHT at the LHC.

In previous works on studying the phenomenology of the LHT \cite{Hubisz:2004ft}, 
the effects of T-odd $SU(2)$ doublet fermions were not included.
A preliminary study on the phenomenology of
these T-odd $SU(2)$-doublet fermions in the LHT was reported in
Ref.~\cite{sasha_pheno}. Although, motivated by the dark matter consideration, 
Ref.~\cite{Chen:2006ie} studied some  interesting processes which include the 
effects of T-odd $SU(2)$ doublet fermions, a complete study 
on the phenomenology of these T-odd fermions in the LHT has not yet been presented.

In this paper, we first stress the important role of the T-odd fermions in
high energy  scattering processes relevant to the LHC, such as
$q\bar{q}\rightarrow W_H^+ W_H^-$, where $W_H^\pm$ is the T-parity partner
of  $W$-boson. We show that it is necessary to include   the contribution
from the t-channel process, via the exchange of these T-odd heavy fermions,
to render its scattering amplitude with a good high energy behavior, so that its partial-wave 
amplitudes respect the unitarity
condition. We also show that its numerical effect cannot be ignored for
studying the collider phenomenology at the LHC.  Furthermore, since the current
experimental constraints of the four-fermion contact interactions place an
upper bound on the T-odd $SU(2)$  doublet fermion masses
\cite{Hubisz:2005tx},  we find not only that the T-odd fermion contribution
to $q\bar{q}\rightarrow W_H^+ W_H^-$ is quantitatively important, but also
that the direct  pair production rate of the T-odd fermions could be significant
at the LHC. To illustrate this point, we classify all production processes
of the new heavy particles predicted by the LHT, and calculate the
corresponding production cross sections and decay branching ratios, including the
effects induced by the T-odd $SU(2)$ doublet fermions.  The rest of this
paper is organized as follows. In Sec.~II, we briefly review the model we
study here, a Littlest Higgs model with T-parity.  
In Sec.~III, we discuss
the high energy behavior of $u\bar{u}\rightarrow W_H^+ W_H^-$ process to
illustrate the importance of the T-odd $SU(2)$ doublet fermion contribution
to  high energy scattering processes, in order to restore the unitarity of
partial wave amplitudes. 
In Sec.~IV, we show our numerical results of the
phenomenological study on the Littlest Higgs model with T-parity at the LHC
energy. Our conclusion is given in Sec.~V.

%%%%%%%%%%%%%%%%%%%%%%%%%%%%
%%%%%%%%%%%%%%%%%%%%%%%%%%%%

\section{A Littlest Higgs Model with T-Parity}
In this section, we briefly review the Littlest Higgs Model with T-parity studied 
in~\cite{Low:2004xc,Hubisz:2004ft,Hubisz:2005tx}
and present our notation of the model.
The Littlest Higgs  model is based on an $SU(5)/SO(5)$ non-linear
sigma model~\cite{Arkani-Hamed:2002qy}. A vacuum expectation value
(VEV) of an $SU(5)$ symmetric tensor field ($\Sigma_0$) breaks the
$SU(5)$ to $SO(5)$ at the scale $f$ with
%\begin{eqnarray}
\begin{equation}
\Sigma_0 = \left(
\begin{array}{ccccc}
0& 0& 0& 1& 0\\
0& 0& 0& 0& 1\\
0& 0& 1& 0& 0\\
1& 0& 0& 0& 0\\
0& 1& 0& 0& 0
\end{array}
\right).
%\end{eqnarray}
\end{equation}
A subgroup $[SU(2)_1\times U(1)_1]\times [SU(2)_2\times U(1)_2]$
of the $SU(5)$ is gauged, and at the scale $f$ it is broken into
the SM electroweak symmetry $SU(2)_L \times U(1)_Y$. The 14
Nambu-Goldstone bosons $\Pi^a$ associated with the global symmetry breaking
decompose under $SU(2)_L \times U(1)_Y$ as ${\bf 1_0}\oplus{\bf
3_{0}}\oplus {\bf 2_{1/2}}\oplus{\bf 3_{1}}$ and they are
parametrized by the non-linear sigma model field $\Sigma =\xi^2
\Sigma_0$ as the fluctuations around the VEV in the broken
directions, where $\xi=e^{i\Pi^a X^a/f}$ and $X^a$ are the
generators of the broken symmetry. 
The components ${\bf 1_0}$ and ${\bf 3_0}$ in the Nambu-Goldstone
boson multiplet are eaten by the heavy gauge bosons associated with the 
gauge symmetry breaking.
The $SU(2)$ doublet ${\bf 2_{1/2}}$ is considered to be the Higgs doublet.
The doublet ${\bf 2_{1/2}}$ and the triplet ${\bf 3_{1}}$ Higgs bosons remain 
in the low energy effective theory, which are introduced through:
\begin{eqnarray}
\Pi^a X^a &=&
\left(
\begin{array}{ccc}
{\bf 0}_{2\times 2} & \frac{H}{\sqrt{2}} &  \Phi  \\
\frac{H^\dagger}{\sqrt{2}} & 0 & \frac{H^{\rm T}}{\sqrt{2}} \\
\Phi^\dagger & \frac{H^*}{\sqrt{2}} & {\bf 0}_{2\times 2}
\end{array}
\right)~{\rm with}~
H=\left(
\begin{array}{c}
-i\pi^+\\
\frac{h+i\pi^0}{\sqrt{2}}
\end{array}
\right)~{\rm and}~
\Phi=\left(
\begin{array}{cc}
-i\phi^{++} & -i\frac{\phi^+}{\sqrt{2}}\\
-i\frac{\phi^+}{\sqrt{2}} & -i \frac{\phi^0+i\phi^P}{\sqrt{2}} 
\end{array}
\right),
\end{eqnarray}
where ${\bf 0}_{2\times 2}$ is a two by two matrix with zero components and 
the superscript {\rm T} denotes taking transpose. 
Here we only
show the doublet Higgs $H({\bf 2_{1/2}})$, where $\pi^+$ and $\pi^0$ are eaten 
by the SM $W$- and $Z$-bosons,
respectively, and the triplet Higgs $\Phi({\bf 3_{1}})$,
which forms a symmetric tensor with components $\phi^{\pm\pm}$, $\phi^\pm$, $\phi^0$ and 
$\phi^P$~\cite{Han:2003wu}. 
Since the non-linear sigma model field $\Sigma$ transforms as $\Sigma\rightarrow V\Sigma V^{\rm T}$
under the $SU(5)$ rotation $V$,
its gauge-invariant kinetic term is
given by
\begin{eqnarray}
{\cal L} &=& \frac{f^2}{8}{\rm Tr}(D_\mu \Sigma)^\dagger (D^\mu \Sigma),
\label{kinetic_sigma}
\end{eqnarray}
where the covariant derivative $D_\mu$ for the $[SU(2)_1\times U(1)_1]\times [SU(2)_2\times U(1)_2]$
gauge symmetry is defined as
\begin{eqnarray}
D_\mu \Sigma &=\partial_\mu\Sigma -i& \sum_{A=1,2}\left[
\bar{g}_A (W_{A \mu}^a Q_A^a \Sigma +\Sigma Q_A^a W_{A \mu}^a)
+\bar{g}'_A(B_{A \mu} Y_A \Sigma +\Sigma Y_A B_{A \mu})\right].
\end{eqnarray}
Here $W_{A\mu}^a$ and $B_{A\mu}$ ($A=1,2$) are gauge bosons, and  
$\bar{g}_A$ and $\bar{g}'_A$ are gauge couplings for $SU(2)_A$ and $U(1)_A$ gauge symmetries, 
respectively. They are related to the SM gauge couplings $g$ for $SU(2)_L$ and $g'$ for $U(1)_Y$ as 
$1/g^2=1/{\bar{g}^2_1}+1/{\bar{g}^2_2}$ and $1/g'^2=1/{\bar{g}'^2_1}+1/{\bar{g}'^2_2}$.
The generators for $SU(2)_A$ (denoted as $Q_A^a$) and for $U(1)_A$ (denoted as $Y_A$) are explicitly 
expressed as 
\begin{eqnarray}
Q_1^a&=&\left(
\begin{array}{ccc}
\sigma^a/2 & {\bf 0}_2 & {\bf 0}_{2\times 2}\\
{\bf 0}_2^{\rm T} & 0   & {\bf 0}_2^{\rm T}\\
{\bf 0}_{2\times 2} & {\bf 0}_2 & {\bf 0}_{2\times 2}
\end{array}
\right),~~~~
Q_2^a=\left(
\begin{array}{ccc}
{\bf 0}_{2\times 2} & {\bf 0}_2 & {\bf 0}_{2\times 2}\\
{\bf 0}_2^{\rm T} & 0   & {\bf 0}_2^{\rm T}\\
{\bf 0}_{2\times 2} & {\bf 0}_2 & -\sigma^{a*}/2
\end{array}
\right),\\
Y_1 &=& {\rm diag.}(3,3,-2,-2,-2)/10,~~~Y_2={\rm diag.}(2,2,2,-3,-3)/10,
\end{eqnarray}
where $\sigma^a$ is the Pauli matrix, ${\bf 0}_2=(0,0)^{\rm T}$ and ``diag.'' denotes
a diagonal matrix.

\subsection{Gauge boson sector}

T-parity~\cite{Cheng:2003ju,Low:2004xc} is naturally introduced in this framework. 
It exchanges $[SU(2)_1\times U(1)_1]$ and $[SU(2)_2\times U(1)_2]$ symmetries.
For example, $W_{1\mu}^a \leftrightarrow W^a_{2\mu}$ and $B_{1\mu}\leftrightarrow B_{2\mu}$ under T-parity.
The Lagrangian Eq.~(\ref{kinetic_sigma}) is invariant under T-parity
if $\bar{g}_1=\bar{g}_2$ and $\bar{g}'_1=\bar{g}'_2$ and $\Sigma$ transforms as
$\Sigma\rightarrow \tilde{\Sigma}=\Sigma_0 \Omega \Sigma^\dagger \Omega \Sigma_0$
with $\Omega ={\rm diag.}(1,1,-1,1,1)$.
Note that the doublet Higgs $H$ (triplet Higgs $\Phi$)
is even (odd) under T-parity. 
The T-even combinations of the gauge fields are SM $SU(2)_L$ gauge
bosons $(W^a_\mu)$ and $U(1)_Y$ hypercharge gauge boson $(B_\mu)$, 
defined as $W_\mu^a=\frac{W_{1 \mu}^a+W_{2\mu}^a}{\sqrt{2}}$ and 
$B_\mu=\frac{B_{1\mu}+B_{2\mu}}{\sqrt{2}}$.
The T-odd combinations are T-parity partners of the SM gauge bosons. 
After taking into account electroweak symmetry breaking, the masses of 
the T-parity partners of
the photon $(A_H)$, $Z$-boson $(Z_H)$ and $W$-boson $(W_H)$ are given by
\begin{eqnarray}
M_{A_H} &=&\frac{g'f}{\sqrt{5}} \left[1-\frac{5v_{SM}^2}{8f^2}+\cdots\right],~
M_{Z_H} \simeq M_{W_H}=gf \left[1-\frac{v_{SM}^2}{8f^2}+\cdots\right].
\end{eqnarray}
Here $v_{SM}$ is the electroweak breaking scale, $v_{SM}\simeq 246$ GeV,
so that at tree level the SM gauge boson
masses can be expressed as $M_W=\frac{g}{2}v_{SM}$ and $M_Z=\frac{\sqrt{g^2+g^{'2}}}{2}v_{SM}$ 
for $W$-boson and $Z$-boson, respectively.
Because of the smallness of $g'$, the T-parity partner of the photon $A_H$ tends to be the
lightest T-odd particle in this framework. Since the lightest T-odd particle is stable, 
it can be an interesting dark matter candidate~\cite{Asano:2006nr}.

Because of the T-parity, SM gauge bosons do not mix with the T-odd heavy gauge bosons
even after the electroweak symmetry breaking.
Consequently, the low energy EW observables are not modified at tree
level.
Since the new heavy T-odd particles always contribute to loops in pairs,
the loop corrections to the EW observables are typically small.
As a result, the new particle mass scale $f$ can be
as low as 500 GeV~\cite{Hubisz:2005tx}, and hence, 
T-odd heavy gauge bosons can be copiously produced at the LHC.

\subsection{T-odd $SU(2)$ doublet fermion sector}

To implement T-parity in the fermion sector, one introduces two $SU(2)$ fermion doublets 
$q_i$ $(i=1,2)$ for each SM fermion doublet~\cite{Low:2004xc,Hubisz:2004ft,Hubisz:2005tx}. 
Here $q_i$ are the doublet under $SU(2)_i$ $(i=1,2)$, 
and T-parity exchanges $q_1$ and $q_2$. The T-even combination of $q_i$ is the SM 
fermion doublet and the other T-odd
combination is its T-parity partner. To generate a heavy mass for the T-odd fermion doublet,
we introduce the following interaction, as suggested in 
Ref.~\cite{Low:2004xc,Hubisz:2004ft,Hubisz:2005tx}:
\begin{eqnarray}
{\cal L}_{\kappa}&=& -\kappa f (\bar{\Psi}_2 \xi \Psi_c
+\bar{\Psi}_1 \Sigma_0 \Omega \xi^\dagger \Omega\Psi_c)+{\rm hermitian~conjugate~(h.c.)}.
\label{kappa}
\end{eqnarray}
Here the fermion $SU(2)$ doublets $q_1$ and $q_2$ are embedded into incomplete $SU(5)$ multiplets
$\Psi_1$ and $\Psi_2$ as
$\Psi_1=(q_1,0,{\bf 0}_2)^{\rm T}$ and $\Psi_2=({\bf 0}_2,0,q_2)^{\rm T}$,
and the doublets $q_1$ and $q_2$
are explicitly written as
$q_A=-\sigma_2 \left(
u_{L_A},
d_{L_A}
\right)^{\rm T}=\left(
i d_{LA},
-i u_{LA}
\right)^{\rm T}$
with $A=1,2$.
Under the global $SU(5)$,
the multiplets $\Psi_1$ and $\Psi_2$ transform as
$\Psi_1 \rightarrow V^* \Psi_1$ and $\Psi_2 \rightarrow V \Psi_2$,
where $V$ is an $SU(5)$ rotation matrix.
A multiplet $\Psi_c$ is also introduced as
$\Psi_c=(q_c,\chi_c,\tilde{q}_c)^{\rm T}$,
which transforms non-linearly under $SU(5)$:
$\Psi_c \rightarrow U \Psi_c$ where
$U$ is an unbroken $SO(5)$ rotation matrix in non-linear representation
of $SU(5)$.
The object $\xi$ and the non-linear sigma model field
$\Sigma~(\equiv \xi^2 \Sigma_0)$
transform like
$\xi \rightarrow V\xi U^\dagger=U\xi \Sigma_0 V^T \Sigma_0$
and $\Sigma \rightarrow V\Sigma V^{\rm T}$, respectively,
under $SU(5)$. T-parity transformation laws are defined as follows:
$\Psi_1 \leftrightarrow -\Sigma_0 \Psi_2$,
$\Psi_c \rightarrow -\Psi_c$, and
$\xi \rightarrow \Omega \xi^\dagger \Omega$.
Thus,
$q_1\leftrightarrow -q_2$  and
$\Sigma \rightarrow \tilde{\Sigma}\equiv
\Sigma_0 \Omega \Sigma^\dagger \Omega \Sigma_0$
under T-parity.
One can verify that the interaction in Eq.~(\ref{kappa}) is invariant under
T-parity.

From the interaction in Eq.~(\ref{kappa}), one can see that the T-odd fermion doublet
$q_-\equiv (q_1+q_2)/\sqrt{2}=(id_{L_-},-iu_{L_-})^{\rm T}$ gets a Dirac
mass, with
$\tilde{q}_c\equiv(id_{R_-},-iu_{R_-})^{\rm T}$, as
\begin{eqnarray}
M_{d_-}&\simeq& \sqrt{2}\kappa f,~~M_{u_-}\simeq \sqrt{2}\kappa f
\left(1-\frac{v_{SM}^2}{8f^2}+\cdots\right).
\label{Todd_mass}
\end{eqnarray}
One may think that assuming a large $\kappa$ value, these T-odd fermions will decouple and
hence we may ignore any effects induced by the T-odd $SU(2)$ doublet fermions. However, as pointed out
in Ref.~\cite{Hubisz:2005tx}, there is non-decoupling effect in some four-fermion operators
whose coefficients become larger as the magnitude of $\kappa$ increases.
%\footnote{ The interaction in Eq.~(\ref{kappa}) also induces
%important effect in the Higgs production via gluon-gluon fusion
%process $(gg\rightarrow h)$ as pointed out in
%Ref.~\cite{Chen:2006cs}.} 
The constraint on the four-fermion
contact interaction contributing to the
$e^+e^-\rightarrow q\bar{q}$ scattering sets an 
important upper
bound on the T-odd fermion masses $M_{q_-}$ as~\cite{Hubisz:2005tx}
\begin{eqnarray}
M_{q_{-}}<4.8 \left(
\frac{f}{1~{\rm TeV}} \right)^2~{\rm TeV}.
\label{T-odd-limit}
\end{eqnarray}
Here we have assumed a universal $\kappa$ value to all T-odd fermion
couplings generated by Eq.~(\ref{kappa})
Therefore, the effect of T-odd fermions to high energy collider
phenomenology may not be negligible, and actually it is
quantitatively important as we will discuss in later sections. 
The interaction terms in 
Eq.~(\ref{kappa}) in general contain 
flavor indices, and large flavor mixings
can cause flavor-changing-neutral-current (FCNC) problem~\cite{Hubisz:2005bd}. 
For simplicity, we assume the flavor diagonal and universal $\kappa$ in this study.

In the multiplet $\Psi_c$, there are other extra T-odd fermions. For those fermions,
we simply assume Dirac masses, as suggested in Ref.~\cite{Cheng:2003ju,Low:2004xc}.
Furthermore, we assume that
their Dirac masses are so large (as large as about 3 TeV) that these extra T-odd
fermions are decoupled, but remains to be small enough 
not to generate the naturalness problem
in the Higgs mass parameter. Thus, in our following analysis, we will not consider any effects induced by
these extra T-odd fermions.

\begin{table}[t]
\begin{center}
\begin{tabular}{|c|c|c|c|c|c|c|c|c|}
\hline
      & $q_1$  & $q_2$  & $U_{L_1}$ & $U_{L_2}$ & $U_{R_1}$ & $U_{R_2}$ & $u_{R_+}$  & $d_{R_+}$\\
\hline
$Y_1$ & $1/30$ & $2/15$ & $8/15$    & $2/15$   & $8/15$ & $2/15$ & $1/3$ & $-1/6$ \\
$Y_2$ & $2/15$ & $1/30$ & $2/15$    & $8/15$   & $2/15$ & $8/15$ & $1/3$ & $-1/6$ \\
\hline
\end{tabular}
\end{center}
\caption{$U(1)_A$ charges $Y_A$ for fermions. The SM hypercharge is given by $Y=Y_1+Y_2$.}
\label{U1_charges}
\end{table}
The $U(1)_A$ charges $Y_A$ for fermions are listed in Table~\ref{U1_charges}.\footnote{
Strictly speaking, these $U(1)_A$ charges $Y_A$ ($A=1,2$) for fermions are defined by a
sum of the $U(1)_A$ charges from the original $SU(5)$ and extra fermion $U(1)$ charges.}
Those charges are determined by the gauge invariance of the Yukawa couplings which
we will discuss later.
In addition to the normal SM gauge interactions,
the T-odd fermions interact with their SM partner fermions and the heavy gauge boson as follows:
\begin{eqnarray}
{\cal L}&=&\frac{g}{\sqrt{2}} W^+_{H\mu}(\bar{u}_{L}\gamma_\mu d_{L_-}
+\bar{u}_{L_-}\gamma_\mu d_{L})+{\rm h.c.} \nonumber 
\\
&+&\sum_{f=u,d}\left[(g c_H T_{3f}+g's_H Y') Z_{H\mu}
+(-g s_H T_{3f}+g' c_H Y') A_{H\mu}\right] \bar{f}_L \gamma_\mu f_{L_-}
+{\rm h.c.},
\label{Todd_fermi_Todd_gauge_int}
\end{eqnarray}
where $Y'=-1/10$, and $s_H~(\equiv\sin\theta_H)$ 
describes the degree of mixing between heavy neutral gauge bosons
with 
$s_H\simeq \frac{gg'}{g^2-g^{'2}/5}\frac{v_{SM}^2}{4f^2}$ and $c_H\equiv\cos\theta_H$.
For clarity, the corresponding Feynman rules are presented in Appendix A.
Through these interactions, the T-odd fermion can contribute to 
heavy gauge boson productions.
Also, it can be directly produced via exchanging light gauge bosons,
heavy gauge bosons, and gluons  
at high energy hadron colliders,
such as the LHC, as we will discuss in the following sections.

\subsection{Yukawa couplings for Top and other fermions}

In order to cancel the large radiative correction to Higgs mass parameter induced by top-quark,
we introduce in the top sector the singlet fields $U_{L_1}$ and
$U_{L_2}$, which are embedded,  
together with the $q_1$ and $q_2$ doublets, into the following
multiplets:
$Q_1=(q_1,U_{L_1},0_2)^{\rm T}$ and $Q_2=(0_2,U_{L_2},q_2)^{\rm T}$.
For the top-Yukawa interaction, one can write down the following T-parity invariant
Lagrangian:~\cite{Low:2004xc,Hubisz:2004ft,Hubisz:2005tx}:
\begin{eqnarray}
{\cal L}_t&=& -\frac{\lambda_1f}{2\sqrt{2}} \epsilon_{ijk} \epsilon_{xy}
\left[(\bar{Q}_1)_i \Sigma_{jx} \Sigma_{ky}-
(\bar{Q}_2 \Sigma_0)_i \tilde{\Sigma}_{jx} \tilde{\Sigma}_{ky}
\right] u_{R_+}
\nonumber \\
&&
-\lambda_2 f (\bar{U}_{L_1} U_{R_1}+\bar{U}_{L_2} U_{R_2}) +{\rm h.c.},
\label{top_yukawa_int}
\end{eqnarray}
where $\epsilon_{ijk}$ and $\epsilon_{xy}$ are antisymmetric tensors,
and $i,~j$ and $k$ run over $1-3$ and $x$ and $y$ over $4-5$.
$u_{R_+}$ and $U_{R_i}$ $(i=1,2)$ are $SU(2)$ singlets.
Under T-parity, these fields transform as
%\begin{eqnarray}
$Q_1 \leftrightarrow -\Sigma_0 Q_2~$,
$U_{R_1}\leftrightarrow -U_{R_2}$ and $u_{R_+}\rightarrow u_{R_+}$.
%\end{eqnarray}
The above Lagrangian contains the following mass terms:
\begin{eqnarray}
{\cal L}_t &\simeq&
-\lambda_1 f \left[
\frac{v_{SM}}{f}\left(1-\frac{v_{SM}^2}{4f^2}+\cdots\right) \bar{u}_{L_+} u_{R_+}
+\left(1-\frac{v_{SM}^2}{2f^2}\right) \bar{U}_{L_+} u_{R_+}\right]
\nonumber
\\
&-&\lambda_2 f \left(\bar{U}_{L_+} U_{R_+}+\bar{U}_{L_-} U_{R_-}
\right)+{\rm h.c.}
\label{top-yukawa}
\end{eqnarray}
Here we have defined the T-parity eigenstates as $q_+\equiv (q_1-q_2)/\sqrt{2}=(id_{L_+},-iu_{L_+})$,
$U_{L_\pm}=\frac{U_1\mp U_2}{\sqrt{2}}$
and $U_{R_\pm}=\frac{U_{R_1}\mp U_{R_2}}{\sqrt{2}}$.
One T-odd Dirac fermion $T_-$ ($T_{-L} \equiv U_{L_-},~T_{-R} \equiv U_{R_-}$)
gets a mass $M_{T_-}=\lambda_2 f$ (cf. Eq.~(\ref{top-yukawa})), and
a T-odd combination of the doublets $q_1$ and $q_2$
obtains a mass from ${\cal L}_\kappa$ (cf. Eq.~(\ref{kappa})).
The left-handed (or right-handed) top quark ($t$) is a linear combination of
$u_{L_+}$ and $U_{L_+}$ (or $u_{R+}$ and $U_{R_+}$), and another independent
linear combination is a heavy T-even partner of the top quark $(T_+)$:
\begin{eqnarray}
\left(
\begin{array}{c}
u_{X_+}\\
U_{X_+}
\end{array}
\right) &=&
\left(
\begin{array}{cc}
c_X & s_X\\
-s_X & c_X
\end{array}
\right)
\left(
\begin{array}{c}
t_X\\
T_{+X}
\end{array}
\right)
,~~(X=L,R),
\end{eqnarray}
where the mixings are approximately expressed by
\begin{eqnarray}
s_L = s_\alpha^2 \frac{v_{SM}}{f}+\cdots,~~
s_R = s_\alpha\left[
1-\frac{c_\alpha^2(c_\alpha^2-s_\alpha^2)}{2}\frac{v_{SM}^2}{f^2}+\cdots)\right],
\label{s_LR}
\end{eqnarray}
with $s_\alpha=\lambda_1/\sqrt{\lambda_1^2+\lambda_2^2}$ and 
$c_\alpha=\lambda_2/\sqrt{\lambda_1^2+\lambda_2^2}$.
The masses of the top quark ($t$) and T-even heavy top quark ($T_+$) are given by
\begin{eqnarray}
M_t &=&\lambda_1 c_\alpha v_{SM}\left[1-\frac{c_\alpha^4+s_\alpha^4}{4}\frac{v_{SM}^2}{f^2}
+\cdots\right],~~~
%\\
%\nonumber
M_{T_+}=\frac{\lambda_1}{s_\alpha} f \left[1-\frac{c_\alpha^2 s_\alpha^2}{2}\frac{v_{SM}^2}{f^2}
+\cdots\right].
\end{eqnarray}
Note that the T-even heavy top ($T_+$) is always heavier than the T-odd heavy top
$(T_-)$ in the effective theory considered here. 
The  Feynman rules
of 
SM and heavy gauge boson interactions in the top sector 
are also summarized  in Appendix A.
We note that the coupling strength of $W^+ \bar{t}b$ is $V_{tb}^{eff}=V_{tb}c_L$
(see Appendix A)
where $V_{tb}$ is the $(t,b)$ element of the Cabibbo-Kobayashi-Maskawa (CKM) matrix. 
For our numerical results shown below,
we have assumed $V_{tb}=1$, so that  $V_{tb}^{eff}=c_L=\sqrt{1-s_L^2}$,
where $s_L$ is given in Eq.~(\ref{s_LR}). Once the $V_{tb}^{eff}$ is measured experimentally, then
the parameter space of the model can be further constrained. In other word, when the parameter 
$s_\alpha$ varies, the effective coupling strength of $W^+ \bar{t}b$ also varies
under our assumption $V_{tb}=1$, so that the single-top production rate at the Tevatron and 
the LHC also
varies. As $s_\alpha\rightarrow 0$, it is approaching to the SM $W^+ \bar{t}b$ coupling strength.

\begin{figure}[t]
\centering
\includegraphics*[width=1\textwidth]{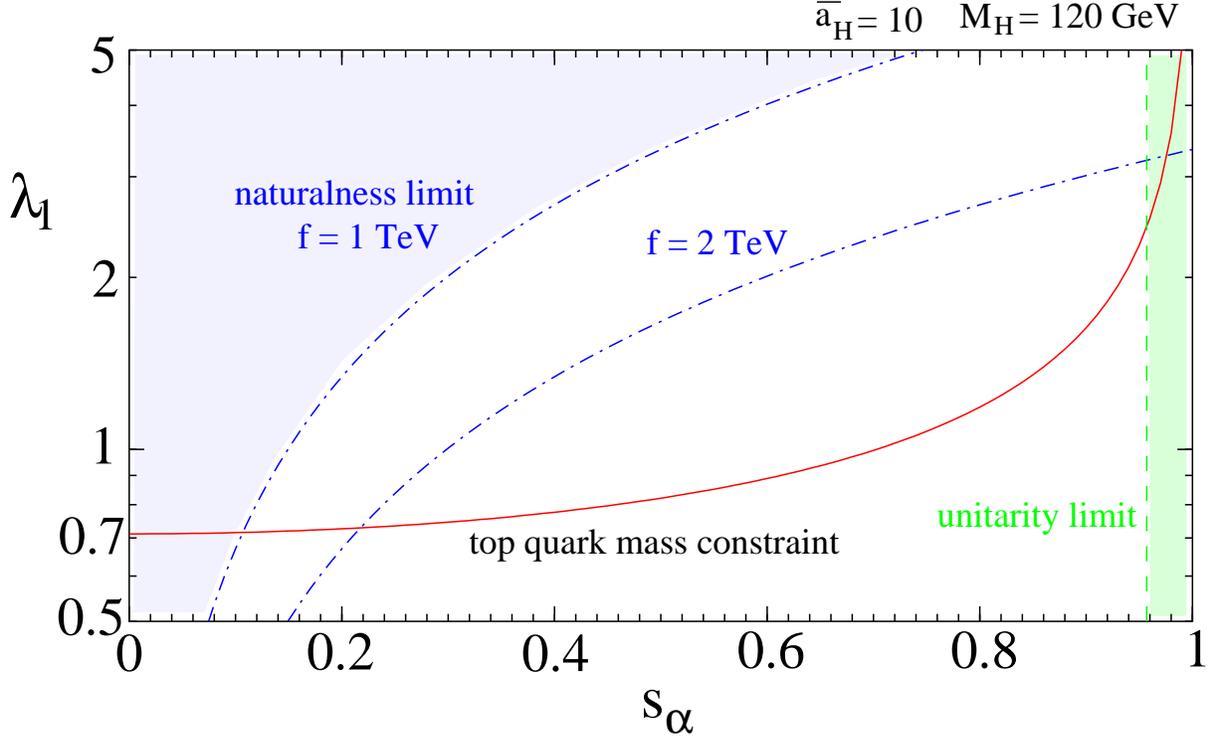}
\caption{Allowed region of parameters $\lambda_1$ and $s_\alpha$.
Solid line (red) represents a relation between
$\lambda_1$ and $s_\alpha$ required by top quark
mass ($M_t=175$ GeV), cf. Eq.~(\ref{lambda1}).
Dashed line (green) shows an upper limit on
$s_\alpha$ from the unitarity bound on the $J=1$
partial wave amplitude in the  coupled system of
$(t\bar{t},~T_+\bar{T}_+,~b\bar{b},~WW,~Zh)$ states,
as expressed in Eq.~(\ref{unitarity}). Dash-dotted
lines (blue) show that naturalness consideration 
puts lower limit on $s_\alpha$ (or equivalently
lower limit on $\lambda_1$), as shown in 
Eq.~(\ref{naturalness}), and the shaded region in
upper-left area of the figure is excluded for $f=1$
TeV. For $f=2$ TeV, the excluded region is extended
to the dash-dotted line with $f=2$ TeV. Here we have
assumed $\bar{a}_H=10$ and $M_H=120$ GeV. }
\label{top_const}
\end{figure}

In the top sector, there are two free parameters $\lambda_1$ and $\lambda_2$, which can be
replaced by $\lambda_1$ and $s_\alpha$ as two independent parameters. 
The experimental value of the top quark mass ($M_t$) gives the relation between 
$\lambda_1$ and $s_\alpha$ as 
\begin{eqnarray}
\lambda_1=\frac{M_t}{v_{SM}}\frac{1}{\sqrt{1-s_\alpha^2}} \geq 0.71
\label{lambda1}
\end{eqnarray}
for $s_\alpha \geq 0$ and $M_t=175$ GeV.
Moreover, following the method presented in Ref.~\cite{Chanowitz:1978uj}, 
we calculated the $J=1$ partial wave amplitudes in the coupled system
of $(t\bar{t},~T_+\bar{T}_+,~b\bar{b},~WW,~Zh)$ states, which are relevant to
the top Yukawa coupling, to estimate the unitarity limit of the
corresponding scattering amplitudes. 
From the unitarity limit,
we can get a mild constraint on the parameters: $s_\alpha/c_\alpha \leq 3.3$, 
which corresponds to
\begin{eqnarray}
s_\alpha\leq 0.96~~{\rm and}~~\lambda_1\leq 2.5,
\label{unitarity}
\end{eqnarray}
cf. Eq.~(\ref{lambda1}). Its detailed discussion is presented in
Appendix B for completeness. 
We could also discuss the ``naturalness'' constraint on these parameters. If we calculate the one-loop
contribution to the Higgs mass parameter ($m_h$) induced by the top sector,
the correction is described by
%\begin{eqnarray}
$\Delta m_h^2 =c \frac{y_t^2}{16 \pi^2} M_{T_+}^2 \equiv a_H M_H^2,$
%\end{eqnarray}
where $y_t=\sqrt{2}M_t/v_{SM}$ and $c$ is a constant of $O(1)$. This correction should not be much 
larger than the Higgs boson (on-shell) mass squared $M_H^2$, otherwise fine-tuning is needed. Thus the coefficient
$a_H$ is a measure of the ``naturalness'' of the Higgs mass correction.
If we take $\bar{a}_H(=a_H/2c)$ to be smaller than $10$, we get the upper
limit on $M_{T_+}$ as
\begin{eqnarray}
M_{T_+}\leq 6.7~{\rm TeV} \sqrt{\frac{\bar{a}_H}{10}}\left(
\frac{M_H}{120~{\rm GeV}}\right) \, .
\end{eqnarray}
In other word, 
\begin{eqnarray}
s_\alpha \geq 0.11 \sqrt{\frac{10}{\bar{a}_H}}\left(
\frac{120~{\rm GeV}}{M_H}\right)\left(\frac{f}{1~{\rm TeV}}\right).
\label{naturalness}
\end{eqnarray}
We summarize these constraints on the parameters of 
the top sector in Fig.~\ref{top_const}.

For the first and second generation up-type quark Yukawa couplings, we 
assume the same forms of Yukawa couplings as those 
for the top quark,  cf. Eq.~(\ref{top-yukawa}),
except that we do not introduce $SU(2)$-singlet fields $U_{L_A}$ and $U_{R_A}$
for the first and second generations because we do not require the cancellation of the
quadratic divergences induced from the light quark sectors, for their Yukawa couplings are tiny.

For down-type quark Yukawa couplings, one of the possible effective
Lagrangians~\cite{Hubisz,Chen:2006cs}\footnote{
We thank J.~Hubisz for pointing out this possibility.
The effective Lagrangian for the down-type quark Yukawa couplings proposed in
Ref.~\cite{Hubisz:2004ft} is not invariant under $U(1)_A$ ($A=1,2$).}
is given by
\begin{eqnarray}
{\cal L}_{\rm down} &=&\frac{i\lambda_d}{2\sqrt{2}}
f \epsilon_{ij} \epsilon_{xyz}\left[
(\bar{\Psi}'_2)_x \Sigma_{i y} \Sigma_{j z} X
-(\bar{\Psi}'_1 \Sigma_0)_x \tilde{\Sigma}_{i y}
\tilde{\Sigma}_{j z} {\tilde{X}}
\right]d_{R_+},
\label{down-yukawa}
\end{eqnarray}
where $\Psi'_1=(-\sigma_2 q_1,0,0_2)^{\rm T}$ and 
$\Psi'_2=(0_2,0,-\sigma_2 q_2)^{\rm T}$.
Here $X$ transforms into $\tilde{X}$ under T-parity, and 
it is a singlet under $SU(2)_i~(i=1-2)$
with its $U(1)_i~(i=1-2)$ charges being $(Y_1,~Y_2)=(1/10,~-1/10)$.
In this paper, we take $X=(\Sigma_{33})^{-1/4}$, where $\Sigma_{33}$ is the 
$(3,3)$ component of the non-linear sigma model field $\Sigma$.

For charged lepton (neutrino) sector, we assume the same Yukawa structure
as that for down-type quark (first and second generation up-type quark) sector.

\subsection{Higgs boson sector}

As we have shown, there are $SU(2)_L$ doublet and triplet Higgs bosons
in the low energy effective theory.
The gauge and Yukawa interactions break the global symmetry, so these Higgs bosons
receive masses from  radiative corrections via  gauge and Yukawa interactions.
Because of the collective symmetry breaking mechanism, the doublet Higgs boson
does not receive large quadratic divergence in its mass parameter, and
hence the natural mass scale of the doublet Higgs boson is of the order of  weak scale.
On the other hand, the triplet Higgs boson mass is not protected by such a  mechanism,
therefore, its mass scale is naturally of the order of $f$. 
Calculating the dominant quadratically divergent top- and gauge-loop corrections 
to the effective Higgs potential,
one gets~\cite{ Arkani-Hamed:2002qy}
\begin{eqnarray}
{\cal L}_{eff}&=&a_t \lambda_1^2 f^4 \epsilon^{wx}\epsilon_{yz} \epsilon^{ijk} \epsilon_{klm}
\left( \Sigma_{iw} \Sigma_{jx} \Sigma^{*my} \Sigma^{*lz}
+\tilde{\Sigma}_{iw} \tilde{\Sigma}_{jx} \tilde{\Sigma}^{*my}\tilde{\Sigma}^{*lz}
\right)\nonumber \\
&+&a_g f^4 \left( g^2 {\rm Tr}\sum_{A=1,2}(Q_A^a\Sigma)(Q_A^a\Sigma)^*
+%\right.\nonumber \\
%&&\left.\hspace{4cm}
+g'^2  {\rm Tr} \sum_{A=1,2}(Y_A\Sigma)(Y_A\Sigma)^*\right),
\\
&\simeq& -M^2_\Phi \left( {\rm Tr}\Phi^\dagger \Phi+\frac{h^4}{16f^2}\right)+\cdots,
\label{phi_mass}
\end{eqnarray}
where $a_t$ and $a_g$ are constants of the order of 1. 
Note that because of the collective symmetry breaking mechanism, the doublet Higgs boson 
does not receive quadratically divergent 
top- and gauge-loop corrections at one-loop level, however, it receives the logarithmically 
divergent one-loop and quadratically divergent two-loop corrections, even though we don't show them
explicitly in Eq.~(\ref{phi_mass}). As shown in Eq.~(\ref{phi_mass}),
the coefficient of the ${\rm Tr}\Phi^\dagger \Phi$ term is
$-M^2_\Phi$.
Hence, the mass of the triplet Higgs boson is
related to the quartic coupling of the doublet Higgs boson. Consequently, 
there is a relation between the triplet and doublet Higgs boson masses, which
is approximately expressed as
\begin{eqnarray}
M_\Phi \simeq \frac{\sqrt{2}M_H}{v_{SM}} f.
\label{triplet_mass}
\end{eqnarray}
In our analysis, we take the doublet
Higgs boson mass $M_H$ as a free parameter, and we calculate the triplet Higgs boson mass
using Eq.~(\ref{triplet_mass})\footnote{ 
After the electroweak symmetry breaking, each component of the triplet
Higgs multiplet receives different correction to their masses, and hence
non-degeneracy will be induced. In our analysis, we have not included
this non-degeneracy in the triplet mass spectrum.
}.
Since the triplet Higgs multiplet
can participate in electroweak interactions and the triplet Higgs boson masses 
can be of the order of TeV, they
can be directly produced at the LHC.

In Table \ref{mass_spectrum}, we list the typical mass spectrum of
the heavy T-parity partners of the SM particles. Here we have taken 
the scale $f$ to be 1 TeV, $s_\alpha=1/\sqrt{2}$ (or, equivalently,
$\lambda_1=\lambda_2\sim {\frac{\sqrt{2} M_t}{v_{SM}}}\sim 1$), $\kappa=1$, and the 
top quark and Higgs boson masses to be $175$ and 120 GeV, respectively.
We will assume this set of model parameter values in the rest of this paper, unless otherwise stated.

\begin{table}[t]
\begin{center}
\begin{tabular}{|c|c|c|c|c|c|c|c|}
\hline
 & $A_H$ & $Z_H~(W_H)$ & $T_+$ & $T_-$ & $u_-$ & $d_-$ & $\Phi$ \\
\hline
Mass (TeV) & 0.15 &0.65 &1.4 &1.0 & 1.4 & 1.4 & 0.69\\
\hline
\end{tabular}
\end{center}
\caption{Typical values of masses for the heavy T-parity partners of the SM particles.
Here we take the scale $f$ to be 1 TeV, $s_\alpha=1/\sqrt{2}$, $\kappa=1$, and the 
top quark and Higgs boson masses to be $175$ and $120$ GeV, respectively. 
}
\label{mass_spectrum}
\end{table}

%%%%%%%%%%%%%%%%%%%%%%%%%%%%%%%%%%%%%%%%%%%%%%%%%%%%%%%%%

%%%%%%%%%%%%%%%%%%%%%%%%%%%%%%%%%%%%%%%%%%%%%%%%%%%%%%%%%

\section{High energy behavior of $u\bar{u}\rightarrow W_{H}^{+}W_{H}^{-}$}
\begin{figure}[t]
\centering
\includegraphics*[width=1\textwidth]{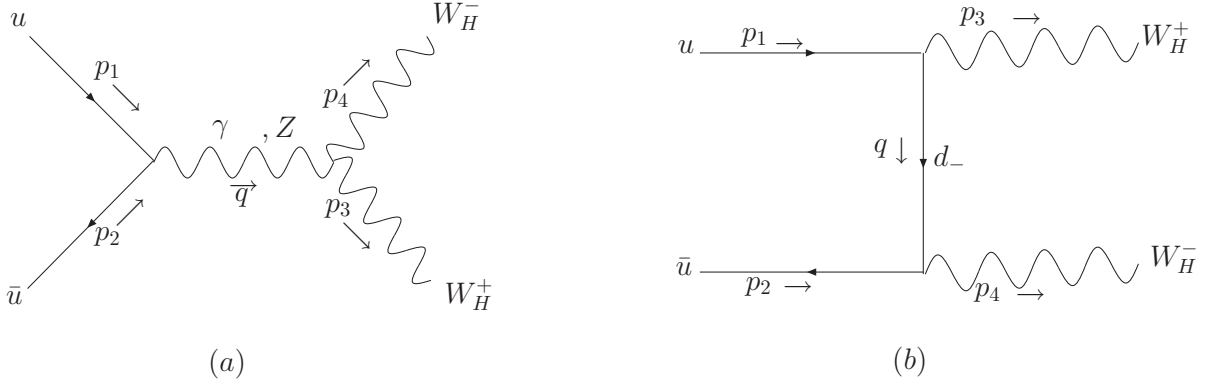}
\caption{Feynman diagrams for $u\bar{u}\rightarrow W_{H}^{+}W_{H}^{-}$.}
\label{uubar-WHWH}
\end{figure}

Before we present a detailed study on the collider phenomenology of the Littlest Higgs model
with T-parity, we stress  in this section 
the importance of the T-odd $SU(2)$ doublet fermion contributions
to high energy processes.
To illustrate the important role of the T-odd $SU(2)$ doublet fermions
in high energy processes,
we discuss the high energy behavior of $u\bar{u} \rightarrow W_H^+ W_H^-$.
The tree-level diagrams for this process are given in Fig.~\ref{uubar-WHWH}. The amplitudes of
the s-channel process with photon and $Z$-boson exchanged are expressed by $A^{\gamma}$ and
$A^{Z}$, respectively, and the amplitude of the t-channel process with
T-odd down-quark $d_{-}$ exchanged is $A^{d_-}$.
%
%\begin{figure}[ht]
For the scattering process $u (p_1) \bar{u} (p_2) \rightarrow W_H^+
(p_3) W_H^- (p_4) $, we find
\begin{eqnarray*}
A^{\gamma} & = & \frac{2e^{2}}{3s}\bar{v}(p_{2})\{(-\not\not{p}_{3}+\not\not{p_{4}})
\varepsilon^{\ast}(p_{3})\cdot\varepsilon^{\ast}(p_{4})
\\
 &  & \,\,\,\,\,\,\,
-2p_{4}\cdot\varepsilon^{\ast}(p_{3})\not{\varepsilon^{\ast}(p_{4})}
+2p_{3}\cdot\varepsilon^{\ast}(p_{4})\not{\varepsilon^{\ast}(p_{3})}\} u(p_{1})
\, ,\\
A^{Z} & = & \frac{e^{2}}{2\sin^{2}\theta_{W}}\frac{1}{s-M_{Z}^{2}}\bar{v}(p_{2})
\{(-\not\not{p_{3}}+\not\not{p_{4}})\varepsilon^{\ast}(p_{3})\cdot\varepsilon^{\ast}(p_{4})\\
 &  & \,\,\,\,\,\,\,-2p_{4}\cdot\varepsilon^{\ast}(p_{3})\not{\varepsilon^{\ast}(p_{4})}+2p_{3}
\cdot\varepsilon^{\ast}(p_{4})\not{\varepsilon^{\ast}(p_{3})}\}(L+R)u(p_{1})
\, ,\\
A^{d_-} & = & -\frac{g^{2}}{2}\frac{1}{t-M_{d_-}^{2}}\bar{v}(p_{2})\not{\varepsilon}^{\ast}(p_{4})
P_{L}(\not\not{t}-M_{d_-})\not{\epsilon^{\ast}(p_{3})}P_{L}u(p_{1}) \, ,
\end{eqnarray*}
 where $L=(1-\frac{4}{3}\sin^{2}\theta_{w})P_{L}$, $R=-\frac{4}{3}\sin^{2}\theta_{W}P_{R}$,
$\theta_{W}$ is the weak mixing angle, and $P_{L}=\frac{1-\gamma_{5}}{2}$ ($P_{R}\
=\frac{1+\gamma_{5}}{2}$)
is the left-handed (right-handed) projection operator.
In the center-of-mass frame of $W_{H}^{+}W_{H}^{-}$,
the 4-momenta of the particles can be chosen to be
\begin{eqnarray*}
p_{1}  &=&  (E,0,0,E),~~
p_{2}  =  (E,0,0,-E),\\
p_{3}  &=&  (E,\, p\sin\theta,0,p\cos\theta),~~
p_{4}  =  (E,-p\sin\theta,0,-p\cos\theta) \, ,
\end{eqnarray*}
where $E$ is the energy of incoming and outgoing particles, $p$
is the momentum of outgoing heavy gauge bosons and $\theta$ is the scattering
angle. In order to check its high energy behavior, we consider 
the case that both the heavy gauge bosons $W_{H}^{+}$ and $W_{H}^{-}$
are longitudinally polarized. Since the incoming fermion $u$
and anti-fermion $\bar{u}$ have opposite helicities, the helicity
amplitudes of s-channel and t-channel processes can be easily found
to be
\begin{eqnarray*}
A^{\gamma}(-+) & = & \frac{8e^{2}Ep(p^{2}-3E^{2})}{3sM_{W_{H}}^{2}}\sin\theta
\, , \\
A^{Z}(-+) & = & (1-\frac{4}{3}s_{W}^{2})\frac{e^{2}}{s_{W}^{2}(s-M_{Z}^{2})}
\frac{2Ep(p^{2}-3E^{2})}{M_{W_{H}}^{2}}\sin\theta
\, , \\
A^{d_-}(-+) & = & \frac{e^{2}}{s_{W}^{2}(t-M_{d_-}^{2})}\frac{E(2E^{3}\cos\theta+p^{3}
-3pE^{2})}{M_{W_{H}}^{2}}\sin\theta
\, ,
\end{eqnarray*}
 where $s_{W}\equiv \sin\theta_{W}$ , $(-+)$ are the helicities of
$(u\,\bar{u})$, the Mandelstam variables $s\equiv(p_{1}+p_{2})^{2}$ and
$t\equiv(p_{1}-p_{3})^{2}$,
and $M_{Z}$ , $M_{W_{H}}$ and $M_{d_-}$ are the
masses of $Z$-boson, heavy $W$-boson and heavy T-odd down-quark,
respectively. As we take the high energy limit, i.e. $\sqrt{s}\gg M_{X}$
($X=Z$, $W_{H}$ and
$d_-$), each amplitude behaves as follows:
\begin{eqnarray}
A^{\gamma}(-+)  &=&  -\frac{s_{W}^{2}\sin\theta}{3f^{2}}s,~~\\
A^{Z}(-+)  &=& -(1-\frac{4}{3}s_{W}^{2})\frac{\sin\theta}{4f^{2}}s,~~
\\
A^{d_-}(-+) &=& \frac{\sin\theta}{4f^{2}}s.
\end{eqnarray}
It is evident that each term diverges as energy goes to infinity, 
but their sum is zero
because of the cancellation between the s-channel and t-channel contributions.
Therefore, we conclude that including the contribution from the T-odd 
down-quark is essential to warrant a good high energy behavior of the 
scattering process $u\bar{u}\rightarrow W_{H}^{+}W_{H}^{-}$.
This can be illustrated by partial-wave analysis, as to be given below.

\begin{figure}[t]
\centering
\includegraphics[width=0.7\textwidth]{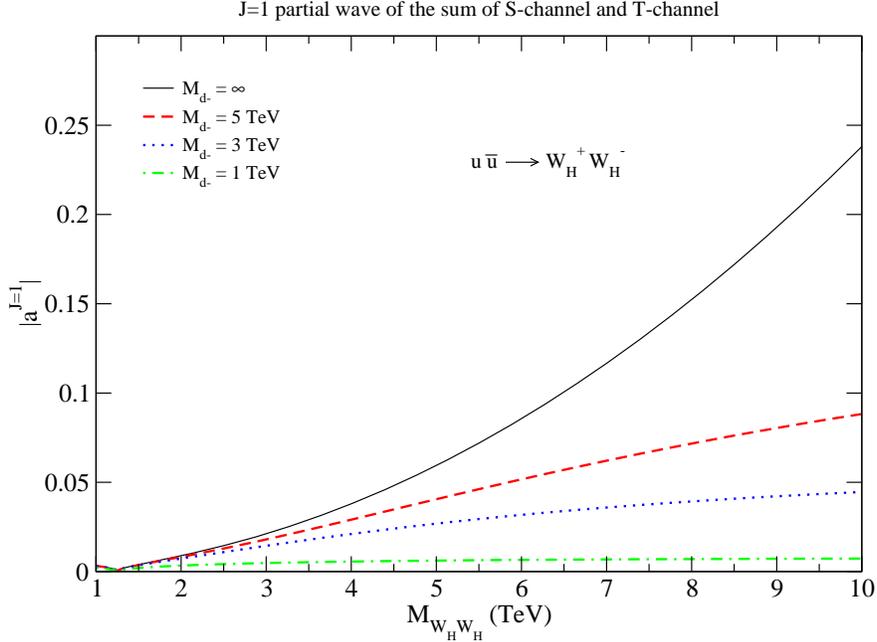}
\caption{$J=1$ partial-wave amplitude of the 
$u\bar{u}\rightarrow W_{H}^{+}W_{H}^{-}$ process,
as a function of $M_{W_H W_H}(=\sqrt{s})$. The
plots are shown for $M_{d_{-}}=1$, 3, 5 TeV and
$\infty$.}
\label{J1amplitude}
\end{figure}

The $J=1$ partial-wave amplitude (denoted as $a^{J=1}$) of
the $u\bar{u}\rightarrow W_{H}^{+}W_{H}^{-}$ process, for producing
longitudinal $W_H$'s, consists of two contributions: one from s-channel,
another from t-channel. We find 
\begin{eqnarray*}
a_{\rm s-channel}^{J=1} & = & \frac{\alpha s}{48\sqrt{2}s_{W}^{2}M_{W_{H}}^{2}}\beta(3-\beta^{2})
\, , \\
a_{\rm t-channel}^{J=1} & = & \frac{\alpha s}{64\sqrt{2}s_{W}^{2}M_{W_{H}}^{2}}
\int_{-1}^{1}\frac{\sin^{2}\theta(2\cos\theta+\beta^{3}-3\beta)}{1-\beta\cos\theta+\frac{2M_{d_-}^{2}}{s}}d\cos\theta
\, ,
\end{eqnarray*}
where $\alpha=\frac{e^{2}}{4\pi}$ and 
$\beta\equiv {\sqrt {1- 4 M_{W_H}^{2}/ s}}$ .
($e$ is the unit of electric charge.)
When $s\gg M_{W_{H}}^{2}$ and $s\gg
M_{d_-}^{2}$, we have 
\begin{equation}
a_{\rm s-channel}^{J=1} = - a_{\rm t-channel}^{J=1} = \frac{\alpha s}{24\sqrt{2}s_{W}^{2}M_{W_{H}}^{2}}
\,.
\end{equation}
In Fig.~\ref{J1amplitude}, we show the $J=1$ partial-wave amplitude of
the $u\bar{u}\rightarrow W_{H}^{+}W_{H}^{-}$ process, as a function of
the invariant mass $(\sqrt{s})$  of the $W_{H}^{+}W_{H}^{-}$  pair, for cases with $M_{d_{-}}=1$,
3, 5 TeV and $\infty$. 
We found  that the unitarity is not
violated up to about 15 TeV in the decoupling limit of the T-odd down
quark, i.e. $M_{d_{-}}\rightarrow \infty$. On the other hand, 
the constraint on the  four-fermi operator 
contributing to the $e^+e^-\rightarrow q\bar{q}$ scattering sets an
important upper limit on the T-odd fermion mass,  as shown in
Eq.~(\ref{T-odd-limit}),  thus the decoupling limit of the T-odd 
fermions is not a realistic assumption and
the T-odd fermion contribution generates an important correction to
$u\bar{u}\rightarrow W_{H}^{+}W_{H}^{-}$ process. This contribution
was not taken into account
in the previous study~\cite{Hubisz:2004ft}.
As we show later, the theoretical prediction of the pair
production cross section  of heavy $W$-boson significantly
depends on the mass of the T-odd down-quark.
%\begin{eqnarray}
%M_{\rm odd}<4.8 \left(
%\frac{f}{1~{\rm TeV}} \right)^2~{\rm TeV}.
%\label{T-odd-limit}
%\end{eqnarray}
%
Moreover, because the mass of the T-odd $SU(2)$ doublet fermions cannot be
too heavy, cf. Eq.~(\ref{T-odd-limit}), they can be copiously
produced at the LHC. Therefore, in the following section, we study the collider
phenomenology of the LHT with emphasis on the contributions of the
T-odd fermion to the production of the heavy T-parity partners
(either bosons or fermions) at the LHC. 

\newpage

%%%%%%%%%%%%%%%%%%%%%%%%%%%%%%%%%%%%%%%%%%%%
%%%%%%%%%%%%%%%%%%%%%%%%%%%%%%%%%%%%%%%%%%%%

\section{Phenomenology}

The Little Higgs mechanism which provides the cancellation of
dominant
quadratic divergences from the top-quark and SM gauge bosons
in Higgs boson mass term, demands the
presence of the partners to SM fermions and bosons. In particular,
detection of the T-even and T-odd partners of top quark would
provide a clear hint for the Little Higgs mechanism with T-parity. 
Similarly, the heavy T-odd gauge
bosons of the electroweak gauge boson partners are also essential
components of the Little Higgs mechanism. 
Many studies~\cite{Hewett:2002px,Burdman:2002ns,Han:2003wu,Perelstein:2003wd,%
Hubisz:2004ft,Han:2005ru,Cheng:2005as,Berger:2005ht,Meade:2006dw} 
have been presented in the literature on the
detection strategies for T-even and/or T-odd partners of top quark 
and  heavy T-odd gauge bosons  at high energy colliders. 
On the other hand, the  T-odd $SU(2)$ partners of SM fermions
of the first and  second generations received so far  little
attention with respect to collider phenomenology.
As we have shown, these fermions are crucial component of  LHT
for providing  its consistency with respect to unitarity 
and viability
with respect to constraints from contact interactions.

In this section, we  first discuss
the production of these heavy T-odd fermions, either produced in pairs
or in association  with heavy T-odd gauge bosons
at the LHC. Then, we discuss
the impact of  T-odd fermion contribution
to the production of heavy T-odd gauge
boson pairs. As discussed in the previous section, it is necessary to
include the contribution from these heavy T-odd fermions to yield an
unitary scattering amplitude. For completeness, we will also discuss the
production of heavy T-odd triplet Higgs bosons.
In the end of this section, we will briefly discuss the potential of the
LHC for testing LHT by classifying several  most interesting 
experimental signatures.

\subsection{Direct Production Rates at the LHC}

Given the model described in section II, we can easily calculate
the direct production rates of non-SM fermions, gauge bosons and
triplet Higgs bosons. In our numerical results, we have used
CTEQ6M parton distribution functions~\cite{Pumplin:2002vw} 
with the renormalization and
factorization scales chosen to be the invariant mass of the
constituent process. Only the leading order results are reported
here. For our phenomenological analysis we 
have implemented the complete LHT  into CalcHEP 
package~\cite{Pukhov:2004ca}\footnote{We are grateful to Alexander Pukhov 
for developing new CalcHEP version while visiting Michigan State University.}
and used it in our analysis. 
To check our analytical derivation of the effective lagrangian
for the implementation of the LHT  into CalcHEP,
we applied LanHEP package~\cite{Semenov:1998eb}
for automatic generations of Feynman rules for the CalcHEP.
Indeed, it turned out that independent implementation
of the model was crucial for the cross check
of the previous studies~\cite{Hubisz:2004ft}.

In our analysis we fix the model parameters to be
$\kappa=1$, $s_{\alpha}=1/\sqrt{2}$
(or equivalently, $\lambda_1=\lambda_2\sim \frac{\sqrt{2} M_t}{v_{SM}}\sim 1$),
the Higgs boson mass $M_H$=120~GeV, and the top-quark mass $M_t=175$~GeV,
while studying the signal production rates as a function of the new particle mass scale $f$.
With this choice of the model parameters, the effective $W^+\bar{t}b$ coupling
is $\frac{g}{\sqrt{2}} V_{tb}^{eff}\gamma_\mu P_L$ with $V_{tb}^{eff}=c_L=\sqrt{1-s_L^2}
\simeq1-0.008\left(\frac{1~{\rm TeV}}{f}\right)^2$.

\subsubsection{Quark-Quark production rates}

\begin{figure}[htb]
\centering
\includegraphics[width=0.4\textwidth]{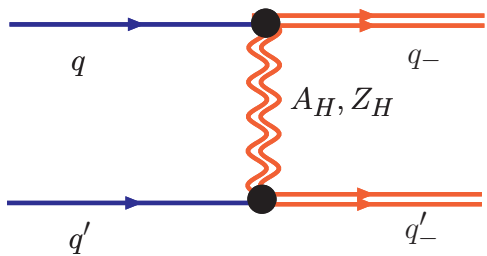}
\caption{Representative Feynman diagram for 
 $pp \rightarrow q_{-} q_{-}^{(')}$ 
via  t-channel exchange 
of T-odd photon $A_H$ and T-odd $Z$-boson $Z_H$.}
\label{qq_EW}
\end{figure}

\begin{figure}[htb]
\centering
\includegraphics[width=0.99\textwidth]{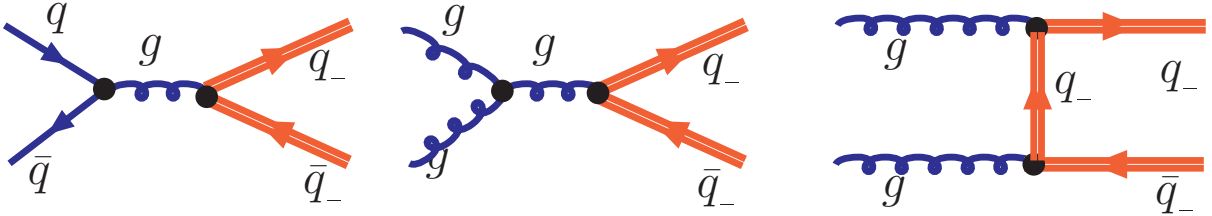}
\caption{QCD Feynman diagrams for 
$pp \rightarrow q_{-} \bar{q}_{-}$ 
process.}
\label{qQ_QCD}
\end{figure}

The LHC is a proton-proton hadron collider, so that a heavy T-odd quark,
denoted as $q_-$ can be copiously produced in pairs as long as its mass is
not too large.  There are two main mechanisms of T-odd quark pair production. 
Firstly, $q_- q_-^{(')}$, same-sign-charge quarks can be produced 
via exchanging the  T-odd heavy photon and $Z$-boson ($A_H$ and $Z_H$)
in $t$~(or $u$)-channel processes initiated by
same-sign-charge light quarks. A respective Feynman diagram corresponding to  this process is
shown in   Fig.~\ref{qq_EW}. Secondly, $q_- \bar{q}^{(')}_-$ pair production  takes place
via both electroweak and obviously dominating QCD processes. The respective
QCD  Feynman diagrams for this process are  shown in   Fig.~\ref{qQ_QCD}.

%%%%%%%%%%%%%%%%%%%%%

%
%%%%%%%%%%%%%%
\begin{figure}[htb]
\centering
%{
%\includegraphics[width=0.5\textwidth]{figs/qq_m34_6m.eps}%
\includegraphics[width=0.8\textwidth]{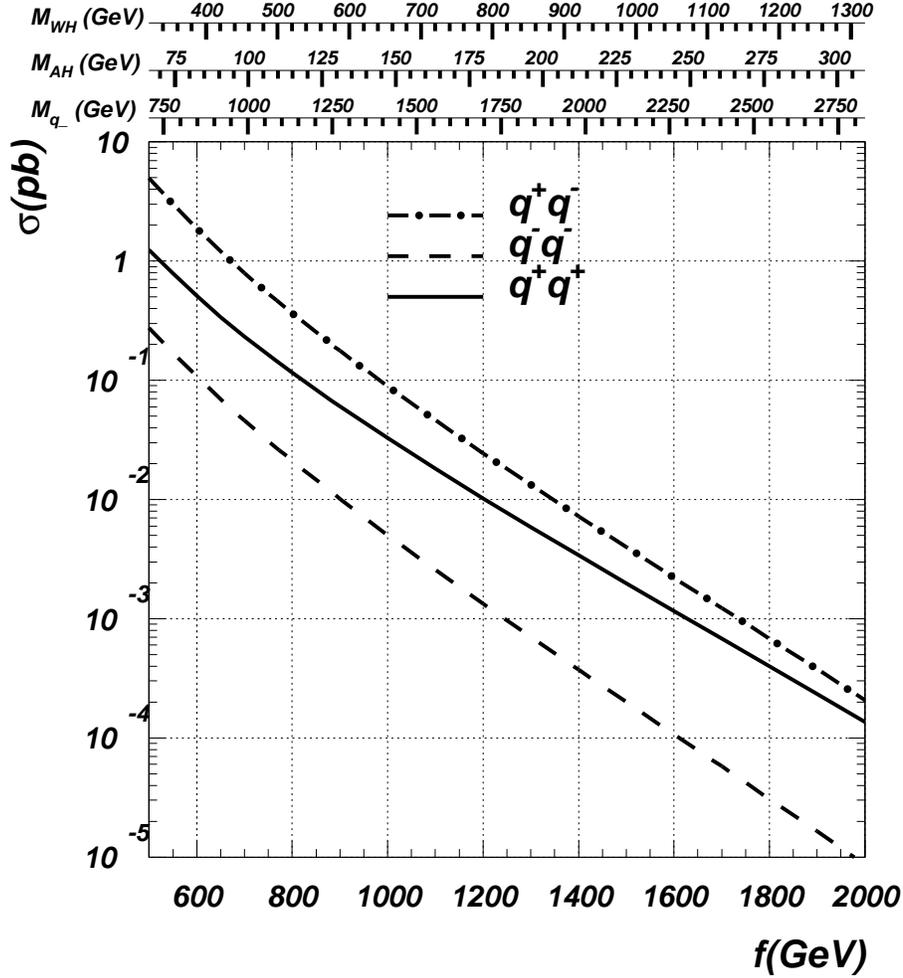}%
\caption{The first and second generation heavy T-odd quark production 
cross sections at the LHC, 
where $q^+ = \lbrace u_-, c_-,{\bar d_-}, {\bar s_-}\rbrace$
and 
$q^- = \lbrace {\bar u_-}, {\bar c_-} , d_-, s_-\rbrace$.
The solid curve presents the production cross section
 of heavy quark pairs with positive
charges ($q_-^+q_-^+$),  dashed curve is for the
production of heavy quark pairs with negative charges 
($q_-^-q_-^-$) and  dot-dashed curve is for the production of heavy quark pairs
with opposite-sign charges ($q_-^+q_-^-$). 
The corresponding masses of the new heavy particles relevant to
the production processes under consideration are listed in the 
top margin of the figure,
corresponding to the respective $f$ values at the bottom.
}
\label{qq}
\end{figure}

In Fig.~\ref{qq} we present pair production rates  of
first and second generation heavy T-odd quarks versus $f$ value,
organized by their electric charges.
The corresponding masses of the new heavy particles relevant to
the production processes under consideration are listed in the 
top margin of the figure,
corresponding to the respective $f$ values at the bottom.
For example, in Fig.~\ref{qq},
for $f=1$~TeV one has
$M_{q_-}\simeq 1.4$~TeV,
$M_{W_H}\simeq 0.65$~TeV and
$M_{A_H}\simeq 0.15$~TeV.

The solid curve presents the production cross section
 of heavy quark pairs with positive
charges, $q_-^+q_-^+$, which includes,  for example, $u_- u_-$, 
${\bar d}_- {\bar d}_-$ and $u_-{\bar d}_-$  pairs. 
The dashed curve is for the
production of heavy quark pairs with negative charges, 
$q_-^-q_-^-$, which includes, for
example, 
${\bar u}_- {\bar u}_-$, 
$d_- d_-$  and $d_-{\bar u}_-$ 
pairs. The dot-dashed curve is for the production of heavy quark pairs
with opposite-sign charges, $q_-^+q_-^-$, which includes, for example, 
${u}_-{d}_-$ and   $\bar{u}_- {\bar d}_-$  pairs. 
It is evident that the heavy T-odd quark pair
production rates are sizable. The production rate of positive charge
pairs is larger than that of the negative charge pairs because of the
larger  parton density associated with positive charge pair production
in proton-proton collision. 
One should notice that electroweak $q_-^+q_-^+$
production is comparable with essentially QCD $q_-^+q_-^-$
production! 
This happens because the production of heavy quark pairs with positive
charges is initiated by both valence quarks in the proton
which have higher parton density than that contributing to
 either QCD or EW $q_-^+q_-^-$ production.
Furthermore, $q_-^+q_-^+$ ($q_-^-q_-^-$) production becomes even more sizable
as compared to $q_-^+q_-^-$ production
when   $f$ (and so the T-odd quark mass) increases, since  
the contribution from valence quarks becomes more
important in the large $x$-value region. 
This is an important result because the  $q_-^+q_-^+$ ($q_-^-q_-^-$) production 
can provide an exciting experimental signatures at the LHC, 
as we shall discuss together with their detection strategies in the following  subsections.

\begin{figure}[htb]
\centering
\includegraphics[width=0.8\textwidth]{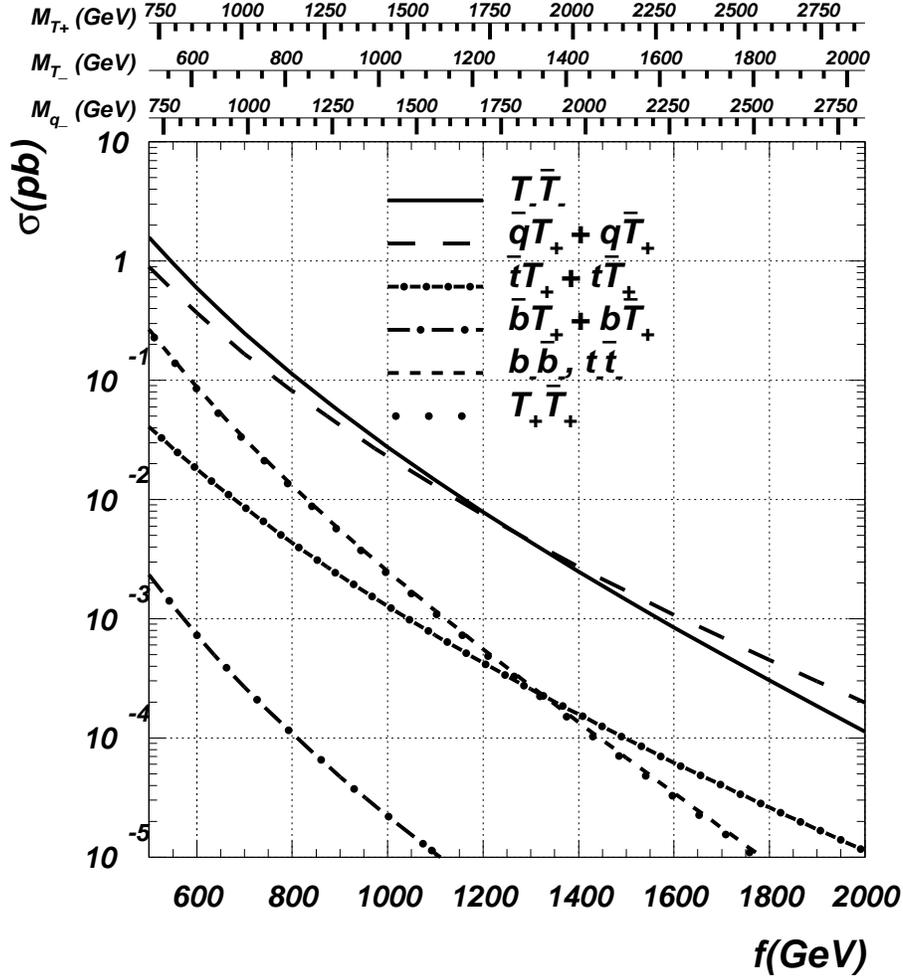}
\caption{The third generation heavy T-odd and T-even quark production 
cross sections at the LHC.\label{q3q3}}

%}
\end{figure}
%%%%%%%%%%%%%%%%%%%%%%%%%%%%%%%%%%%%%%%%%%%%%%%%%%%%%%%%%

In Fig.~\ref{q3q3} we present various production rates of 
heavy T-even and T-odd  top quark pairs as well as the rate of
single T-even heavy top quark associatively produced with SM
light quarks  as a function of $f$. The T-odd bottom quark
pair production rate is also given.

The  T-odd heavy singlet top
quark pairs ($T_- {\bar T}_-$)  have the largest cross section (solid curve)
because in the LHT, considered here,
the  T-odd heavy singlet top quark ($T_-$)
is lighter than the T-even heavy top ($T_+$).
Note that the mass of T-odd doublet quarks
is determined by the choice of $\kappa$ value
which is taken to be 1. In this case
T-odd heavy doublet top quark ($t_-$) mass
is larger than the $T_-$ mass
and is about the same as the $T_+$ mass.

As $f$ increases,
both $T_-$ and $T_+$ become heavier, and the single-$T_+$
production in association with light
quarks ($\bar{q}T_+ + q\bar{T}_+$) (long-dashed curve)
rate becomes larger than the $T_- {\bar T}_-$ rate.
This is because of the phase space suppression in $T_- {\bar T}_-$
(or $T_+ {\bar T}_+$ - dotted curve) 
pair production, for producing two heavy particles, as compared to
producing only one heavy particle in single-$T_+$ event.
Furthermore, the single-$T_+$ production mechanism is dominated by
longitudinal $W$-boson fusion with the incoming bottom quark in
the t-channel production process, similar to the SM t-channel
single-top production~\cite{Yuan:1989tc,single_top_t}.
Due to the collinear enhancement for
the light quark emitting a $W$-boson in the high energy region,
the constituent cross section of single-$T_+$ process does not
drop as fast as that of pair production process.
We note that for a fixed $T_+$ mass, the single-$T_+$ production rate 
is proportional to $s_\alpha^2/c_\alpha^2$.
This is because the coefficient of $W^+ \bar{T}_+b$ coupling is 
$V_{tb}^{eff}\frac{s_L}{c_L}\sim V_{tb}^{eff} s_\alpha^2\frac{v_{SM}}{f}\simeq 
V_{tb}^{eff} \frac{s_\alpha}{c_\alpha}\frac{M_t}{M_{T_+}}$, cf. Appendix A.
In Fig.~\ref{q3q3}, we also show the production rates of the T-odd
$t_-{\bar t}_-$ and $b_- {\bar b}_-$ pairs (short-dashed curve), 
where $t_-$ and $b_-$
are  originated from the T-odd $SU(2)$ doublet quark fields, and their
masses are generated from the $\kappa$ term of the effective
Lagrangian.
One can see that $T_+ {\bar T}_+$ and 
$t_-{\bar t}_-$ (or  $b_- {\bar b}_-$) production cross sections are very close 
to each other because of the same production mechanism and
the  similar masses of $T_+$ and $t_-$ ($b_-$) 
(for this particular choice of model parameters).
Fig.~\ref{q3q3} also presents cross sections for 
the associate $tT_+$ (short dot-dashed line) and $bT_+$ (long dot-dashed line)
productions. The $tT_+$ production rate dominates over
the $bT_+$ rate because the diagram with t-channel $W$-boson exchange
plays the leading role for the $tT_+$ production, and the
similar diagram for $bT_+$ production is suppressed by CKM
matrix elements.
For example, it is suppressed by $V_{cb}$
in the $cb\to b T_+$ production process.

%\newpage
\subsubsection{Quark-Boson production rates}
%%%%%%%%%%%%%%%%%%%%%%%%%%%%%%%%%%%%%%%%%%%%%%%%%%%%%%%%%
\begin{figure}[htb]
\includegraphics[width=0.8\textwidth]{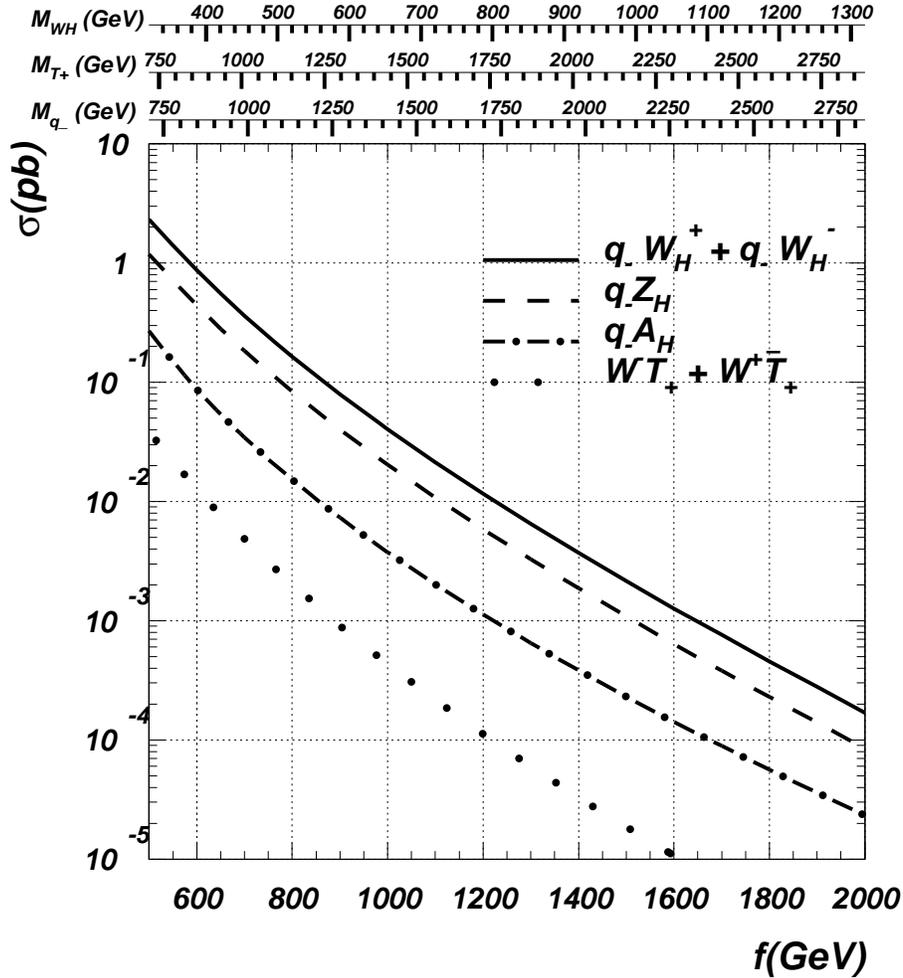}
\caption{Heavy quark-boson associated production rates at the LHC.}
\label{vp}
%%%%%%%%%%%%%%%%%%%%%%%%%%%%%%%%%%%%%%%%%%%%%%%%%%%%%%%%%
\end{figure}

Another production mechanism for heavy T-odd quarks in the LHT
is via associated production with heavy T-odd gauge bosons.
Since the initial state of the scattering processes is T-even, the
final state has to be a pair of T-odd particles. For example, the
$d_- W^+_H$ pair can be produced via the 
$u g\to d_- W^+_H$
production.
In Fig.~\ref{vp}, we show the
associated production rates of heavy T-odd gauge bosons
with all possible T-odd heavy quarks and anti-quarks, including
the T-odd partner of heavy top (anti-)~quark,
as a function of $f$ value.
The fractional contribution from $t_{-}W^{-}_{H}$ 
(and ${\bar t}_{-}W^{+}_{H}$),
initiated by an incoming $b$-quark, to the $q_{-}W^{-}_{H}$ production
is at the percent level, because of the smallness of $b$-quark parton
density inside the proton.
Similarly, the fractional contribution from $b_{-}Z_{H}$ 
(and ${\bar b}_{-}Z_{H}$) to the $q_{-}Z_{H}$ production
is also at the percent level, while the $t_{-}Z_{H}$ contribution is not
included because we do not take top quark as a parton in our calculation.
The same conclusion for $q_{-}Z_{H}$ production also holds for $q_{-}A_{H}$
production after substituting $Z_{H}$ by $A_{H}$.

One can see that $q_-W_H$ (solid line) associate production
is the dominant one, $q_-Z_H$ (dashed line) production rate is about a factor of 2
smaller due to the ratio in their couplings $|g_{qq_-W_H}/g_{qq_-Z_H}|\simeq\sqrt{2}$,
and $q_-A_H$ (dot-dashed line) production is suppressed even more
due to $|g_{qq_-W_H}/g_{qq_-A_H}|\simeq 5\sqrt{2} \cot\theta_W$.

We note that due to T-parity, the
T-even heavy top quark $T_+$ can be produced associatively with
the SM (hence, T-even) gauge bosons, not the T-odd heavy gauge
bosons, whose production rates are also given in Fig.~\ref{vp}
(dotted line). Since $bT_+W$ coupling is suppressed as $v/f$,
one can see that $T_+W^-$  (${\bar T}_+ W^+$) rate
is significantly smaller than the $q_-W_H$ rate,
and this suppression 
obviously grows with the increase of $f$ value.

\subsubsection{Boson-Boson production rates}
\begin{figure}[t]
\includegraphics[width=0.8\textwidth]{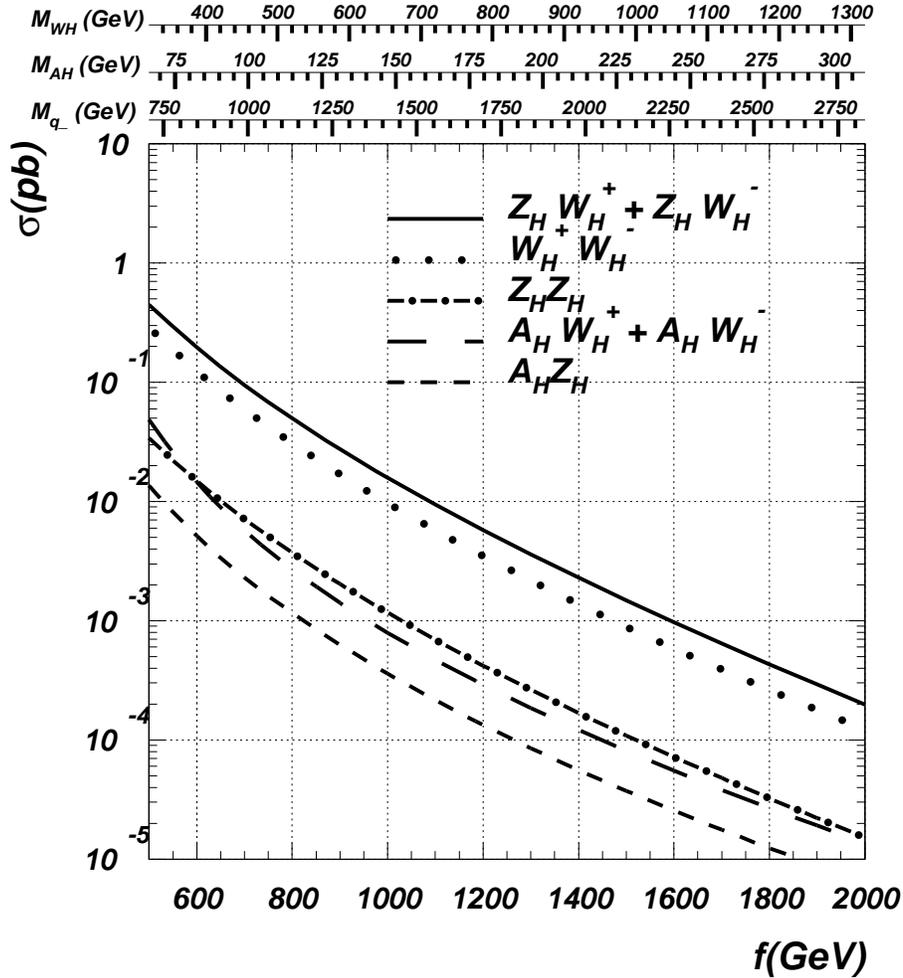}
\caption{Heavy T-odd gauge boson pair production rates  at the LHC.}
\label{vv}
\end{figure}

As discussed in the previous section, the presence of the T-odd
heavy quarks in the model is essential for unitarising the
scattering amplitudes of $qq \to V_H V_H$ processes, where $V_H$
denotes T-odd heavy electroweak gauge bosons. In Fig.~\ref{vv} we
show all possible T-odd heavy gauge boson pair production cross sections
versus $f$ values.
We note that due to the destructive effect from the t-channel
T-odd heavy quark exchange diagram, which is needed to respect  unitarity in high
energy region, the predicted T-odd gauge boson pair production
rates are smaller than those reported in 
Ref.~\cite{Hubisz:2004ft} where the important T-odd heavy quark exchange
diagram was not included in the calculations.

%%%%%%%%%%%%%%%%%%%%%%%%%%%%%%%%%%%%%%%%%%%%%%%%%%%%%%%%%
%\begin{figure}[htb]
%%%%%%%%%%%%%%%%%%%%%%%%%%%%%%%%%%%%%%%%%%%%%%%%%%%%%%%%%

\begin{figure}[htb]
\includegraphics[width=0.8\textwidth]{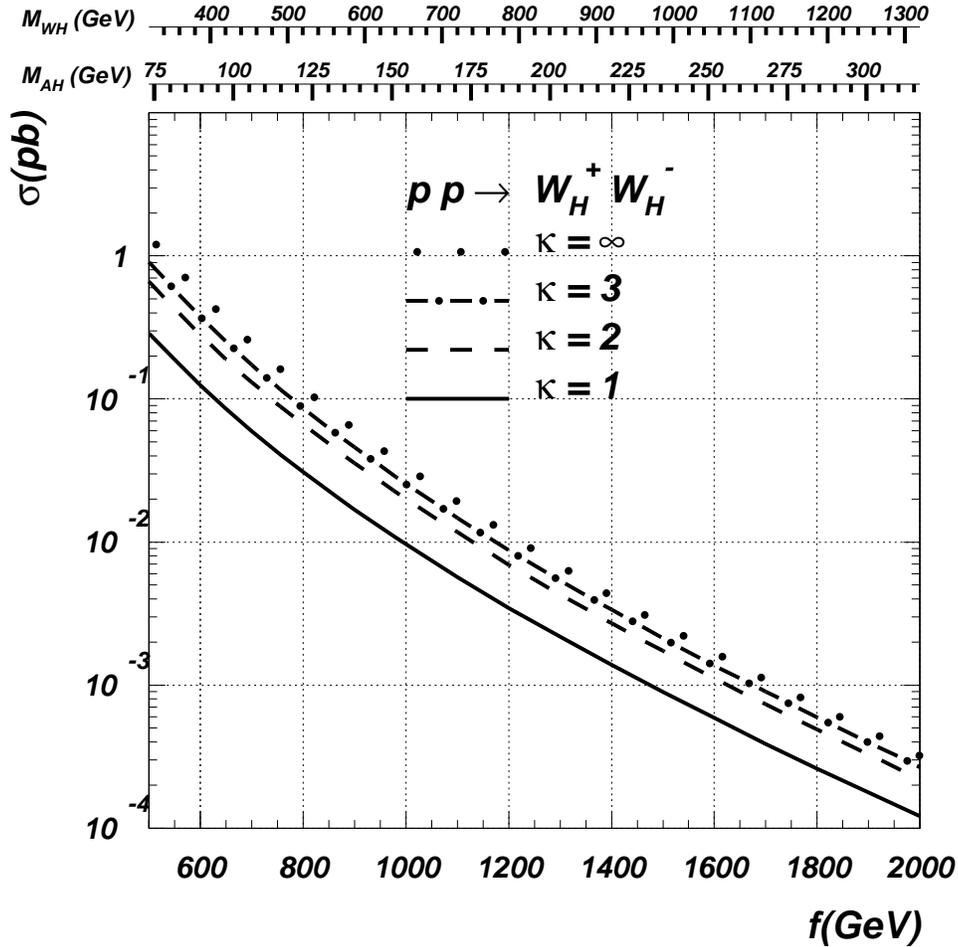}
\caption{Heavy T-odd gauge boson pair, $pp\to W_H^{+} W_H^{-}$, production rates at the LHC.}
\label{whwhkappa}
\end{figure}

Moreover, it is not a constant suppression factor in every
production channel such that the relative difference between the
$Z_H W^\pm_H$ and $W^+_H W^-_H$ rates is much smaller than that reported in
Ref.~\cite{Hubisz:2004ft}. To examine the dependence on model parameters, 
we show in Fig.~\ref{whwhkappa} 
 the production cross section 
of $W^+_H W^-_H$ pair at
the LHC as a function of $f$ for various choices of $\kappa$
values. We note that the curve for $\kappa \to \infty$
corresponds to the calculation without including the T-odd heavy
quark contribution which overestimates $W^+_H W^-_H$ production rate
by a significant factor.
In the later section, we shall come back to discuss its detection
strategies at the LHC.

%\newpage
\subsubsection{Higgs-Higgs production rates}
%%%%%%%%%%%%%%%%%%%%%%%%%%%%%%%%%%%%%%%%%%%%%%%%%%%%%%%%%
\begin{figure}[htb]
\includegraphics[width=0.8\textwidth]{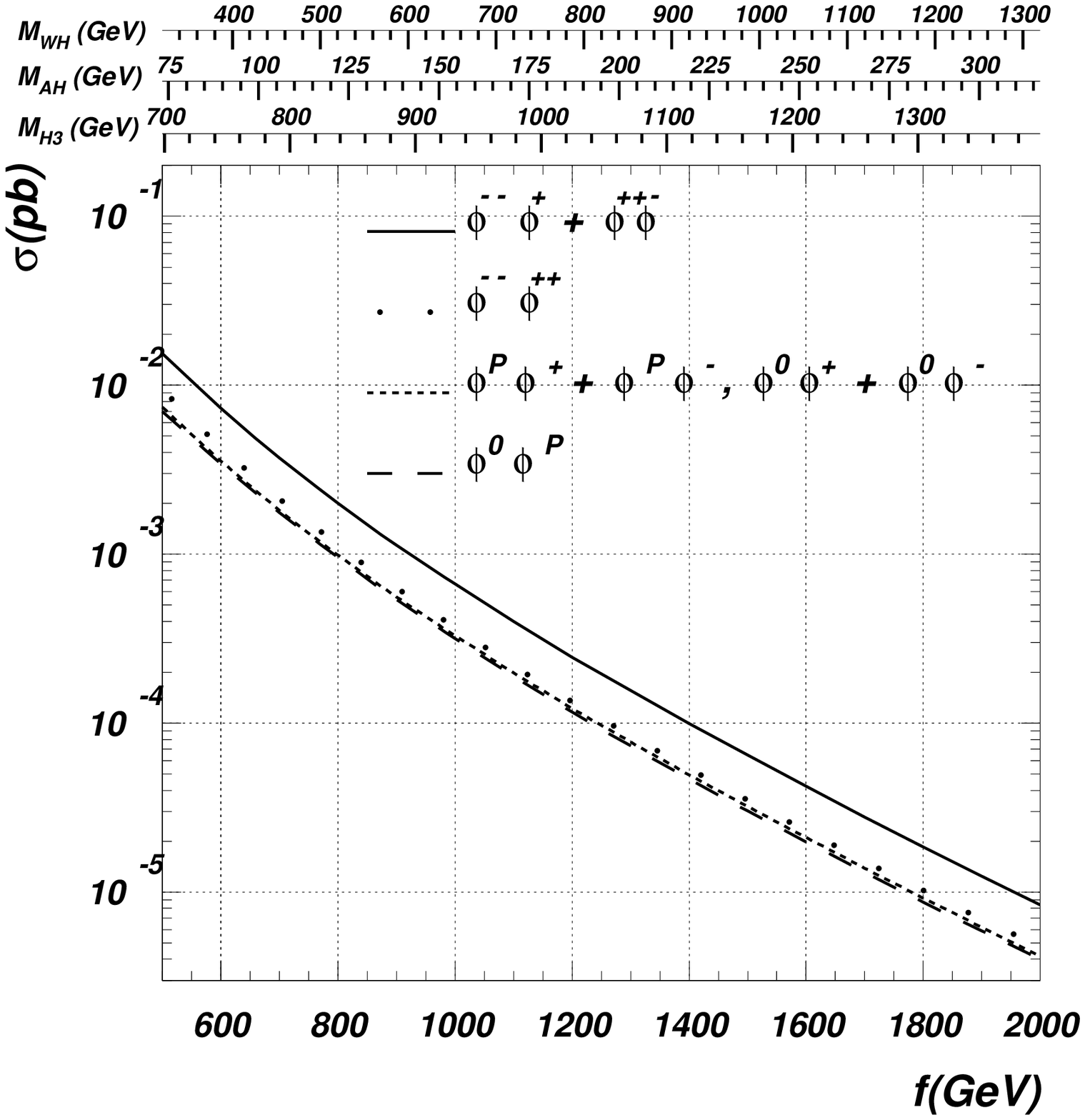}
\caption{Heavy T-odd Higgs production rates at the LHC.}
\label{phiphi}
\end{figure}
%%%%%%%%%%%%%%%%%%%%%%%%%%%%%%%%%%%%%%%%%%%%%%%%%%%%%%%%%

In the LHT, the direct production mechanism of the normal
(T-even) Higgs boson is similar to the SM Higgs boson production
though with somewhat suppressed couplings. We refer the reader to
Ref.~\cite{Chen:2006cs} for more detailed discussions. In high energy 
collision,
the T-odd triplet Higgs bosons can be produced in
$qq \to \phi \phi$ processes at the tree level via gauge interactions of $\phi$, where
$\phi$ denotes any of the T-odd heavy triplet Higgs bosons. Their
direct production rates are small because at tree level they are
produced via s-channel processes 
with highly virtual gauge boson
propagators. Though t-channel diagrams also take place, 
they are strongly suppressed because they 
involve heavy T-odd quarks 
and the  $qq_-\phi$ coupling is suppressed at least by  $v/f$.

Nevertheless, the T-even Higgs bosons can be copiously produced
from the decay of T-odd heavy quarks, as to be discussed  below. 
For that, we shall first examine the decay
branching ratios of the T-odd heavy quarks and gauge bosons
predicted in this model.

\def\baselinestretch{0.9}
%\newpage

\subsection{Decay branching ratios}

%\begin{table}
%
%\begin{center}
% use packages: array

%\end{center}

{
\begin{table}
 \begin{tabular}{|c|c|c|c|c|c|}
\hline
Particle & Decay Mode & Branching Ratio(\%) & Particle & Decay Mode & Branching Ratio(\%) \\
\hline 
$u_{-}$ & $W_{H}^{+}$ $d$ & 61  & $d_{-}$ & $W_{H}^{-}$ $u$ & 62  \\ 
        & $Z_{H}$ $u$     & 30  &         & $Z_{H}$ $d$     & 31  \\ 
        & $A_{H}$ $u$     & 8.6 &         & $A_{H}$ $d$     & 6.3 \\ 
\hline
$b_{-}$ & $W_{H}^{-}$ $t$ & 60  & $t_{-}$ & $W_{H}^{+}$ $b$ & 62  \\ 
        & $Z_{H}$ $b$     & 32  &         & $Z_{H}$ $t$     & 29  \\ 
        & $A_{H}$ $b$     & 6.6 &         & $A_{H}$ $t$     & 8.2 \\ 
\hline
$T_{-}$ & $A_{H}$ $t$     & 100 & $T_{+}$ & $W^{+}$ $b$ & 46  \\
        &                 &     &         & $Z$ $t$     & 22  \\
        &                 &     &         & $H$ $t$         & 20  \\
        &                 &     &         & $A_{H}$ $T_{-}$     & 12  \\
\hline
$W_{H}^{+}$ & $A_{H}$ $W^{+}$ & 100 & $Z_{H}$ & $A_{H}$ $H$ & 100  \\
\hline
$\phi^{+}$ & $A_{H}$ $W^{+}$ & 100 & $\phi^{0}$ & $A_{H}$ $Z$ & 100  \\
\hline
$\phi^{p}$ & $A_{H}$ $H$    & 100  &            &             & \\
\hline
\end{tabular}
\caption{Decay branching ratio of heavy 
particles in Littlest Higgs Model with T-parity. 
Values in this table are calculated with parameters 
$\kappa = 1$, $f = 1$ TeV, $M_{H} = 120$ GeV 
and $M_t = 175$ GeV. We notice that 
for this set of model parameter values,
the 
triplet Higgs $\phi^{++}$ doesn't have two-body decay modes at tree level.}
\label{table}
\end{table}

In order to study the phenomenology of the T-odd heavy particles
predicted in the LHT, we need to know about their decay
branching ratios. In addition to the SM parameters, the dominant
two-body decay modes of the first and second generation T-odd
quarks only depend on two more parameters: $f$ and $\kappa$,  i.e.
$f$ determines the mass of the T-odd heavy gauge bosons and both $f$ and 
$\kappa$ determine the mass of T-odd heavy quarks. If $\kappa$ is
of the order of 1, then because of the smallness of gauge coupling
strength, the T-odd gauge bosons are typically lighter than the
T-odd quarks. When the lightest T-odd particle (LTP) is $A_H$ so that it
could be a good candidate for dark matters, the heavy T-odd quarks
mainly decay into a normal QCD jet plus a T-odd heavy gauge boson
$W^\pm_H$, $Z_H$ or $A_H$. As shown in Table~\ref{table}, the decay
branching ratio (BR) into $W^\pm_H + jet$ is about twice of
BR($Z_H + jet$) and one order of magnitude larger than BR($A_H +
jet$) for $f=1$\,TeV and $\kappa=1$. This feature also holds for
the T-odd heavy top ($t_-$) and bottom ($b_-$) quarks which are
originated from the T-odd $SU(2)$ doublet quark fields and gain their masses from
$\kappa$ terms. The T-odd heavy $SU(2)$ singlet top quark ($T_-$),
originated from the top quark Yukawa interaction Lagrangian,
decays almost 100\% into the $t A_H$ mode. The T-even heavy $SU(2)$
singlet top quark ($T_+$) has a more complicated decay pattern
and can decay into $W^+ b$, $H t$, $Z t$ and $A_H t_-$ modes
with nontrivial dependence on the model parameters such as $f$,
$\lambda_1$ and $\lambda_2$ (or, equivalently, the masses of heavy
T-odd gauge bosons, $T_+$ and $T_-$). 
In Fig.~\ref{tpdecay}, 
we present the decay
branching ratios for the above decay channels of $T_+$ as a
function of $c_\alpha$ (left frame) for $f$=1~TeV,
and as a function of $f$ for $s_\alpha=1/\sqrt{2}$ (right frame)
\footnote{ We have found  disagreement for branching 
ratios of $T_+$ for  small values of $s_\alpha$
as compared to results reported in Ref.~\cite{Hubisz:2004ft}.
In our paper, $c_\alpha$ corresponds to $s_\lambda$ 
of Ref.~\cite{Hubisz:2004ft}.
For the sake of comparison
we present $c_\alpha$ dependence of BR$(T_+)$
in the left frame of Fig.~\ref{tpdecay}.}.
One can see that at $c_\alpha\simeq 1$, 
BR$(T_+\to Ht)$ becomes dominant since for small $s_\alpha$,
 $HT_+t$ coupling is proportional to  $c_\alpha$
 while the couplings of $T_+$ in the other decay channels  are suppressed
 by $s_\alpha$. 
Note that for our analysis, the coefficient of the $W^+\bar{t}b$ coupling
$V_{tb}^{eff}\equiv V_{tb}c_L$ varies as $c_L$, cf. Appendix A.
Here $V_{tb}$ is taken to be 1.
On the other hand,
the BR of T-odd heavy quarks are quite
insensitive to the LHT  parameters as long as the mass of the
T-odd heavy quark is larger than $A_H$. For example, the values of
BRs shown in Table~\ref{table} also hold 
(within a few percents) for $f=0.5-1$\,TeV range.
Hereafter, we will take $f=1$~TeV as the reference point.

\begin{figure}[htb]
\includegraphics[width=0.99\textwidth]{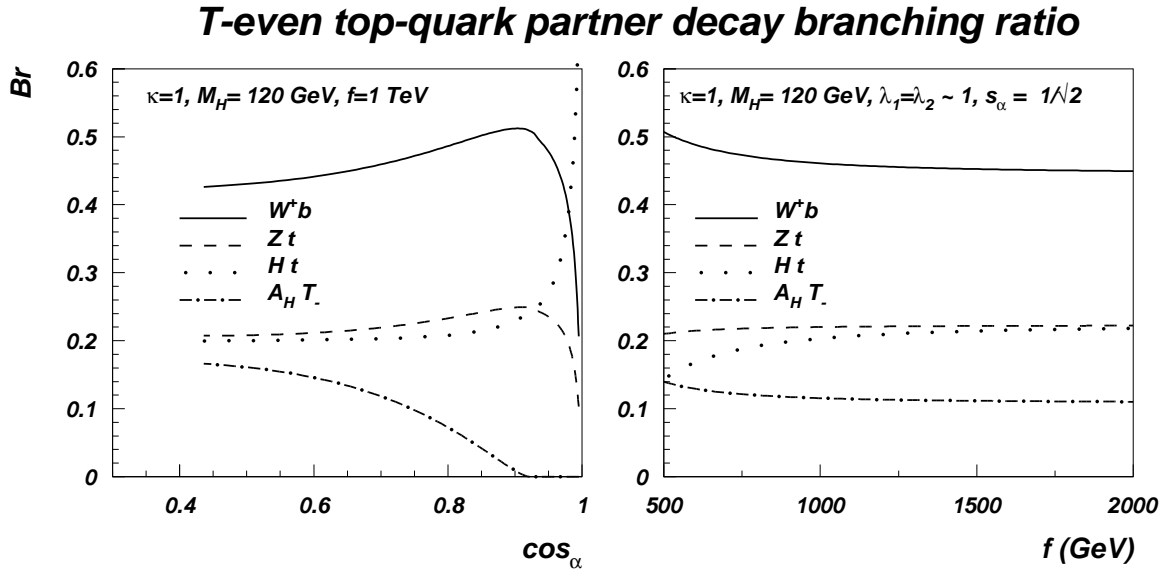}
\caption{T-even heavy top decay branching 
ratios for $\kappa=1$, $\lambda_1=\lambda_2$~(or, $s_\alpha=1/\sqrt{2}$) and  $M_H=120$~GeV.}
\label{tpdecay}
\end{figure}

The striking feature of the T-odd heavy gauge boson decay pattern
is that $W^\pm_H$ almost exclusively decay into a $W^\pm A_H$
pair, while $Z_H$ decays into a $Z H$ pair, for $\kappa$ being of
the order 1 and the mass of the (T-even) Higgs boson is about
120\,GeV. This is because the masses of $W^\pm_H$ and $Z_H$ are
about the same and are smaller than the T-odd heavy quark masses
(unless $\kappa$ is much less than 1). 
In such cases, the normal
(T-even) Higgs boson can be copiously produced from the decay of
T-odd heavy gauge boson $Z_H$ which can be produced either
associatively with T-odd heavy quarks or another heavy T-odd gauge
bosons, as discussed above.

For the chosen model parameters, with $\kappa = 1$ and
$\lambda_1 = \lambda_2$~(or, $s_\alpha=1/\sqrt{2}$) 
and $M_H=120$~GeV, there is no tree-level two-body decay mode for 
the T-odd doubly charged triplet Higgs boson, $\phi^{\pm\pm}$,
while $\phi^{\pm}$ decays into  $W^{\pm}  A_H$ mode,
and  $\phi^0$ and $\phi^{P}$ decay into $ZA_H$ and $HA_H$ modes,
respectively. However, for  $M_H\ge 130$~GeV, the $W^\pm_H W^\pm$
mode could be opened for  $\phi^{\pm\pm}$ Higgs boson.

\def\baselinestretch{0.8}
%\newpage

\subsection{Signal processes and the collider signatures}

In this section we shall discuss various experimental signatures
of signal processes at the LHC for the same values of model
parameters as given in the previous section. For simplicity, we
shall concentrate on the pure leptonic decay modes of gauge bosons
in the final decay chain of T-odd heavy quarks and gauge bosons.

\subsubsection{The 1st and 2nd generation heavy T-odd quark pair production}

According to the multiplicity and charge of the leptons produced
from the first and second generation  T-odd heavy quark chain
decays, we can classify the T-odd heavy quark pair event signature
as signal events with like-sign di-leptons, opposite-sign
di-leptons, and single charged lepton with large missing
transverse momentum.

\begin{figure}[htb]
\centering{
\includegraphics[width=0.8\textwidth]{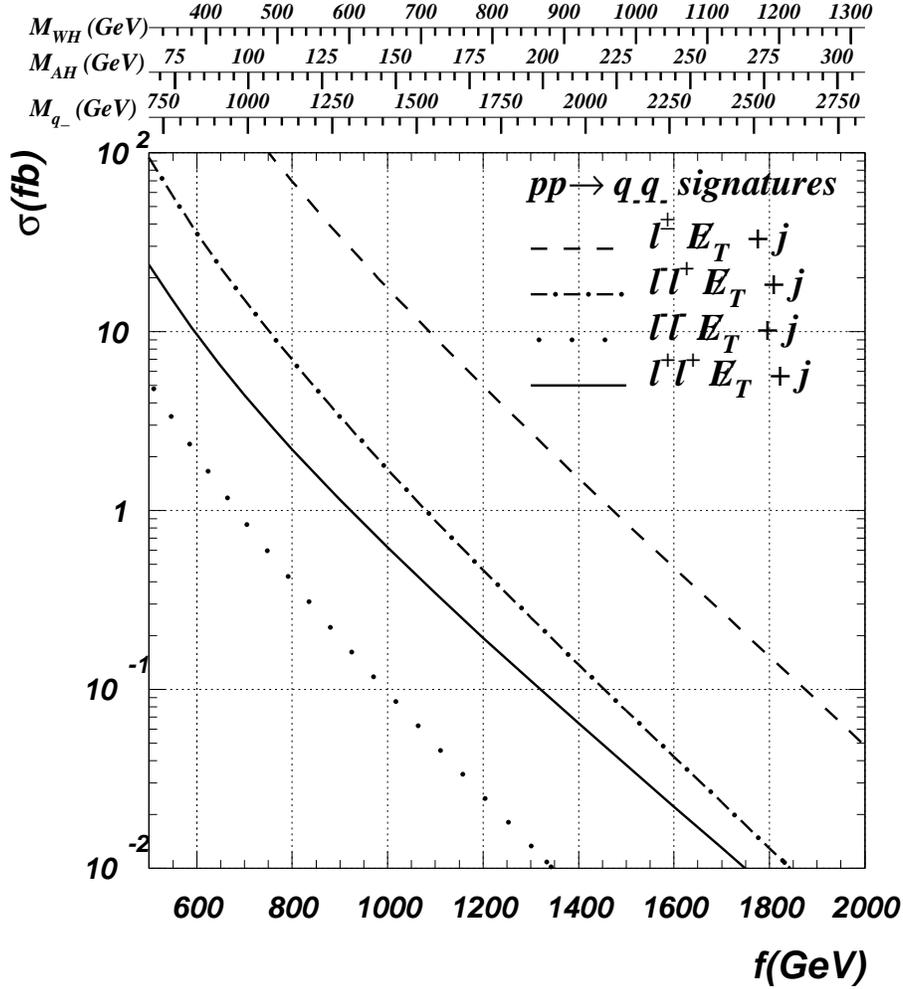}}%
\caption{\label{qqsign}
            Rates for like-sign di-lepton, opposite-sign di-lepton and  
            single charged lepton signatures from 
	    the 1st and 2nd generation heavy T-odd quark pair production at the LHC.}
\end{figure}

%
%\begin{itemize}
\noindent
 {\bf 
 1) like-sign di-lepton ($\ell^\pm\ell^\pm + \etmiss +jets$) signature (LSL)}\\
As shown in Fig.~\ref{qq_EW}, the valence quark initiated $pp\to q_-
q_-$  processes via the
exchange of heavy electroweak gauge bosons could give rise to a
large production rate of signal events with a pair of like-sign
charged leptons in the final state to yield a distinct
experimental signature. For example,
\\
$u_{-}u_{-}      \to  W_H^+ d  W_H^+ d       \to W^+ W^+ A_H A_H d
d $ \ \ \ and \ \ \
$u_{-}\bar{d}_{-}\to  W_H^+ d  W_H^+ \bar{u} \to W^+ W^+ A_H A_H d\bar{u}$\\
chains lead to the $\ell^+\ell^+ + \etmiss +jets$ signature, while
\\
$d_{-}d_{-}  \to  W_H^- u  W_H^-      u  \to W^- W^- A_H A_H uu$
\ \ \ and \ \ \ $d_{-}\bar{u}_{-}\to  W_H^- u  W_H^- \bar{d} \to
W^- W^- A_H A_H u\bar{d}$
\\
processes
produce  $\ell^-\ell^- + \etmiss +jets$ final state.
\\
The overall decay branching ratios for the above reactions can be
easily calculated from Table~\ref{table} which yields $Br[q_{-}q_{-}\to
LSL]=0.62^2 \times (2/9)^2 \simeq 0.019$. Depending on the
values of $f$, the LSL signal event rate  for positively charged leptons
is about at 23 fb level
for a lower value of $f=0.5$~TeV and 
about  0.6 fb  for $f=1$~TeV,
as shown in Fig.~\ref{qqsign}.
LSL  signal event rate for  negatively charged leptons
is 5 fb  and 0.1 fb, respectively, as shown by dotted line 
in Fig.~\ref{qqsign}.

With the high luminosity option of the LHC, around
$300\,{\rm fb}^{-1}$, there will be a large number of signal
events with like-sign di-leptons, with large transverse momentum
($P_T$), and large missing transverse momentum ($\etmiss$)
in the
 $\ell^-\ell^- + \etmiss +jets$
 or
 $\ell^+\ell^+ + \etmiss +jets$ signature. 
The prominent
feature of this signal signature is that it is free of large
$t\bar{t}$ background. This is similar to the case for studying
the longitudinal weak boson scattering processes in the TeV region,
with emphasis on the so called Gold-platted purely leptonic decay
mode of weak bosons. As shown in Ref.~\cite{Bagger:1993zf}, 
after imposing the
kinematic cuts on the charged leptons, the SM background rate,
which is dominated by the intrinsic electroweak $q q W^\pm W^\pm$
production and the $Wt {\bar t}$ associate production, is already
down to the level of a few tenth fb. It is expected that one can
further discriminate the signal event from the SM background event
by requiring a large scalar sum of the transverse momenta, contributed
by the two high $P_T$ charged leptons, jets and $\etmiss$, which
is known as the $H_T$ parameter in the search for top quark at the
Tevatron~\cite{top-observ}. This is because in the signal event, two heavy
T-odd quarks are produced so that the center-of-mass system has a
much larger mass. Furthermore, one can use the kinematic
constraints, similar to those used in the $t {\bar t}$ analysis
carried out at the Tevatron, to purify the data sample with
T-odd heavy quarks. Finally, one can construct the transverse mass
of the final state system, in analogy to the one introduced in
Ref.~\cite{Bagger:1993zf} 
for studying the longitudinal weak boson scattering, to
further discriminate the SM background from the signal events.
Therefore, the LSL signature of the T-odd quark pair events is
expected to provide a clear verification or disproof of the Littlest Higgs model
with T-parity
unless the signal
production rate is largely suppressed for very large
$f$ and therefore very heavy T-odd quarks.
%
%\item
\\
{\bf 2) opposite-sign lepton  
($\ell^\pm\ell^\mp + \etmiss +jets$) signature (OSL)}\\
A shown in Fig.~\ref{qq}, the production of T-odd heavy quark pairs
with opposite electric charges has a higher rate than the
like-sign heavy quark pairs. For example,
\\
 $u_{-}\bar{u}_{-} \to  W_H^+ d  W_H^- \bar{d} \to W^+ W^- A_H A_H \bar{d} d $,\\
\ \ \ \ 
$u_{-}  d_-     \to  W_H^+ d W_H^-        u       \to W^+ W^- A_H A_H du$,
\\
$d_{-}\bar{d}_{-}   \to  W_H^- u  W_H^+   \bar{u}     \to W^- W^+A_H A_H u\bar{u}$, \\
 $d_{-}\bar{u}_{-}\to  W_H^- u  W_H^+ \bar{d} \to W^- W^- A_H A_H u\bar{d}$
\\
processes all give rise to the $\ell^+\ell^- + \etmiss +jets$
signature. 
When the mass of the T-odd heavy quarks increases, the
electroweak production rate becomes more important than the QCD
production rate. One of the reasons is that the former process is
dominated by the t-channel  exchange of a relative
light $A_H$ boson, and the later process is induced by the s-channel 
exchange of a virtual gluon. 
Another reason is that the
former process can be initiated by two valence quarks (via t-channel process) 
while the later process must involve a sea-quark parton 
whose density function becomes smaller in the large $x$-region
for producing a heavier
T-odd heavy quark pair. 

The overall decay branching ratio for the above
reactions is equal to Br[$q_{-}q_{-}\to LSL]$. 
Hence, the OSL
signal event rate is larger than the LSL signal event rate
as indicated by the dot-dashed line in Fig.~\ref{qqsign}.
However, the  OSL signal suffers from a much larger SM background rate
induced by the $t {\bar t}$ production. Nevertheless, the same
strategies discussed above to suppress the SM background rate in
the LSL analysis also applies to the OSL case because the signal
events are all generated from a system with a much larger mass
(i.e., the invariant mass of the heavy T-odd quark pair) as
compared to the SM background processes. To be certain, a detailed
Monte Carlo analysis is needed which is beyond the scope of this
work.
\\
{\bf 3)  Single charged lepton ($\ell^\pm + \etmiss +jets$) signature (1L)}\\
One may also consider the signal event signature with only one
charged lepton in its final state, with one of $W^\pm$ decaying
leptonically and another hadronically. The overall decay branching
for the above reactions is equal to $ Br[q_{-}q_{-}\to 1L]= 
Br[q_{-}q_{-}\to W_{H}W_{H}qq\to 1L]+Br[q_{-}q_{-}\to W_{H}A_{H}qq\to 1L]=
0.62^2
\times 2/9 \times 2/3 \times 2 + 0.62\times 0.086 \times 2/9 \times 2 
\simeq 0.14\simeq 6\times Br[q_{-}q_{-}\to LSL]$. The production rate is also higher,
as presented by the dashed line in Fig.~\ref{qqsign}, for all the above T-odd heavy
quark pair production channels are combined.
On the other hand, the expected background will also be orders of
magnitude higher. Hence, it is more challenging to detect the
signal events in the single charged lepton mode.

\subsubsection{The third generation heavy quark pair production}

In order to cancel the quadratic divergence induced by the top
quark loop for Higgs boson mass correction at the one-loop order,
we need to introduce additional  heavy quarks  (heavy partners of top quark) 
into the LHT
model. 
In general, there are $T_+$, $T_-$,
originated from the top quark Yukawa sector, cf.~Eq.~(\ref{top_yukawa_int}), 
and $t_-$, originated
from the $\kappa$ term interaction with $b_-$ as its isospin
partner, cf.~Eq.~(\ref{kappa}).

\begin{figure}[htb]
\centering{
\includegraphics[width=0.8\textwidth]{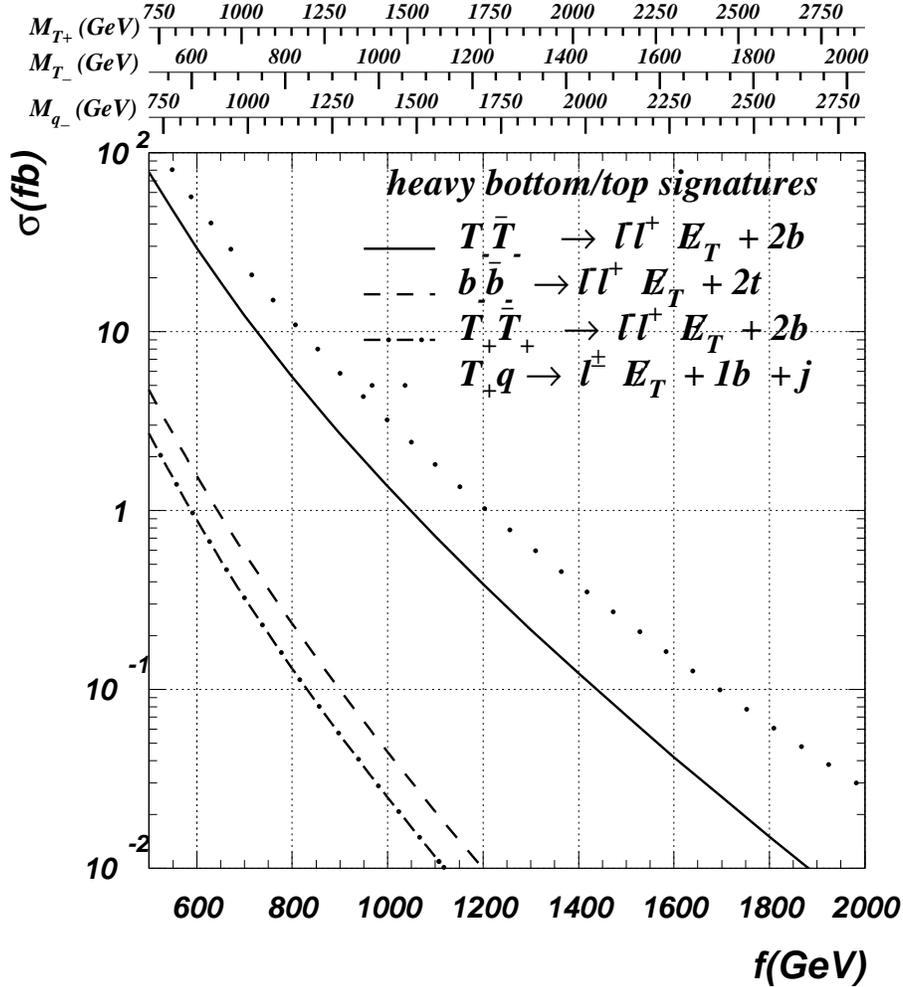}}
\caption{\label{q3q3sign}
            Rates for opposite-sign di-lepton and  
            single charged lepton signatures from 
	    the third generation heavy quark pair production
            at the LHC.}
\end{figure}

 \noindent
 {\bf 1) $T_{-}\bar{T}_{-}$ production}
with
$T_{-}\bar{T}_{-}\to  A_H A_H t\bar{t}$\\
The $T_{-}\bar{T}_{-}$ production rate at the LHC is quite large,
which is about  $30$~fb for $f=1$\,TeV. The experimental
signature of this signal event can be either OSL or 1L. Its
production rate only depends on $T_{-}$ mass  and the decay branching ratio
of $T_{-}\to t A_H$  is about 100\%. Therefore, it is important to
test this production mode at the LHC, for the signal rate can be
predicted with great confidence. We present OSL  rates for
$T_{-}\bar{T}_{-}$ production in Fig.~\ref{q3q3sign}  as solid
line.
There have been a few studies in the literature to
discuss how to detect this channel at the 
LHC~\cite{Hubisz:2004ft,Cheng:2005as},
though more
detailed Monte Carlo analysis is needed to confirm how well this
channel can be detected. It was also pointed out that it could be
very challenging to distinguish this production channel with the
top-squark (stop) pair productions predicted by the Minimal
Supersymmetric Standard Model (MSSM) with the subsequent decay of stop
into top quark and the lightest supersymmetric particle
(neutralino)~\cite{Cheng:2005as}. 
Needless to say that distinguishing  the LHT from the
MSSM generally requires studying of all detectable
experimental  signatures induced by various
production mechanisms predicted by the models.
\\
{\bf 2) $t_{-}\bar{t}_-$ and  $b_{-}\bar{b}_-$ production}\\
For the particular choice of $\kappa=1$, which makes $t_-$ and  $b_-$
heavier then ${T}_{-}$,
the $t_{-}\bar{t}_-$ production rate is at least one  order  of
magnitude (depending on the value of $f$) lower than the $T_{-}\bar{T}_{-}$ rate. 
In case of
$t_{-}\bar{t}_-$ production, there will be two $b$-jets
associatively produced with a pair of OSL or 1L in its event
signature.
%({\bf or} $t{\bar t}$ + Higgs pair + $\etmiss$)  
Likewise, the $b_{-}\bar{b}_-$ process gives rise to a
$t\bar{t}$ pair in addition to the OSL or 1L signature.
%({\bf or} two $b$-jets + Higgs pair + $\etmiss$) 
The rate for  OSL+$t\bar{t}$ signature
is presented in  Fig.~\ref{q3q3sign} 
by dashed line. 
The rate for  OSL+$b\bar{b}$ from $t_-\bar{t}_-$ production
is very similar and is not shown.
Depending on $\kappa$,  $t_{-}\bar{t}_-$ 
production rate could be higher or lower than the  $T_{-}\bar{T}_{-}$ production rate,
making it, respectively,  harder or easier  to observe.
\\
{\bf 3) $T_{+}\bar{T}_{+}$  production}\\
Since ${T_{+}}$ is heavier than $T_{-}$, the  $pp\to
T_{+}\bar{T}_{+}$ production rate 
(similar to the $t_{-}\bar{t}_-$ or  $b_{-}\bar{b}_-$ production rate) 
is at least one  order  of
magnitude 
lower than the  $T_{-}\bar{T}_{-}$ production rate 
 (depending on the value of $f$).
The highest rates are for $T_{+}\bar{T}_{+}\to W^+W^-b\bar{b}$ signature which
should be checked against the  SM $t\bar{t}$ background. 
The rate for  OSL+$b\bar{b}$  signature
 is presented in  Fig.~\ref{q3q3sign} 
by the dot-dashed  line. 
Again,
the techniques discussed about for using the large invariant mass
of the heavy system in the signal event to distinguish it from the
SM background event could be useful for detecting the signal event
in this channel.
\\
{\bf 4) single $T_{+}$  production}\\
 The rate of single-$T_{+}$ production associated with a light quark
 via t-channel electroweak  interaction is
actually higher than the rate of $T_{+}\bar{T}_{+}$ pair
production via strong interaction, as clearly shown in  Fig.~\ref{q3q3}.
The dominant experimental signature of the signal event is the same as the SM
single-top event though it is expected with a much larger missing
transverse momentum. 
In Fig~\ref{q3q3sign} the dotted line presents the rate of 1L signature
originated from the  single $T_{+}$  production in association with the light quark.
Furthermore, the transverse mass of the
signal event will be larger than that of the SM single-top event.
In analogy to the SM single-top event, the single-$T_{+}q$ event
is also characterized by a forward-jet which populates in the
large rapidity region and can be used to suppress $t {\bar t}$ and
$Wb {\bar b}$ backgrounds \cite{Yuan:1989tc}.
 Again, a Monte Carlo study is
needed to draw any definite conclusion about its detection at the
LHC.

\subsubsection{ $q_{-}V_H$ associate  production}

As discussed  in Ref.~\cite{Chen:2006cs}
 Higgs boson production rate via gluon-gluon fusion process
is always smaller than  that predicted by SM.
However, because in most part of the model parameter space, a
 heavy T-odd $Z_H$ decays almost entirely into a $ZH$ pair, it
provides new production channels for the SM-like Higgs boson. The
experimental signature of the $ q_{-}V_H$ pair production can be
classified as follows.
\begin{figure}[t]
\centering{
\includegraphics[width=0.8\textwidth]{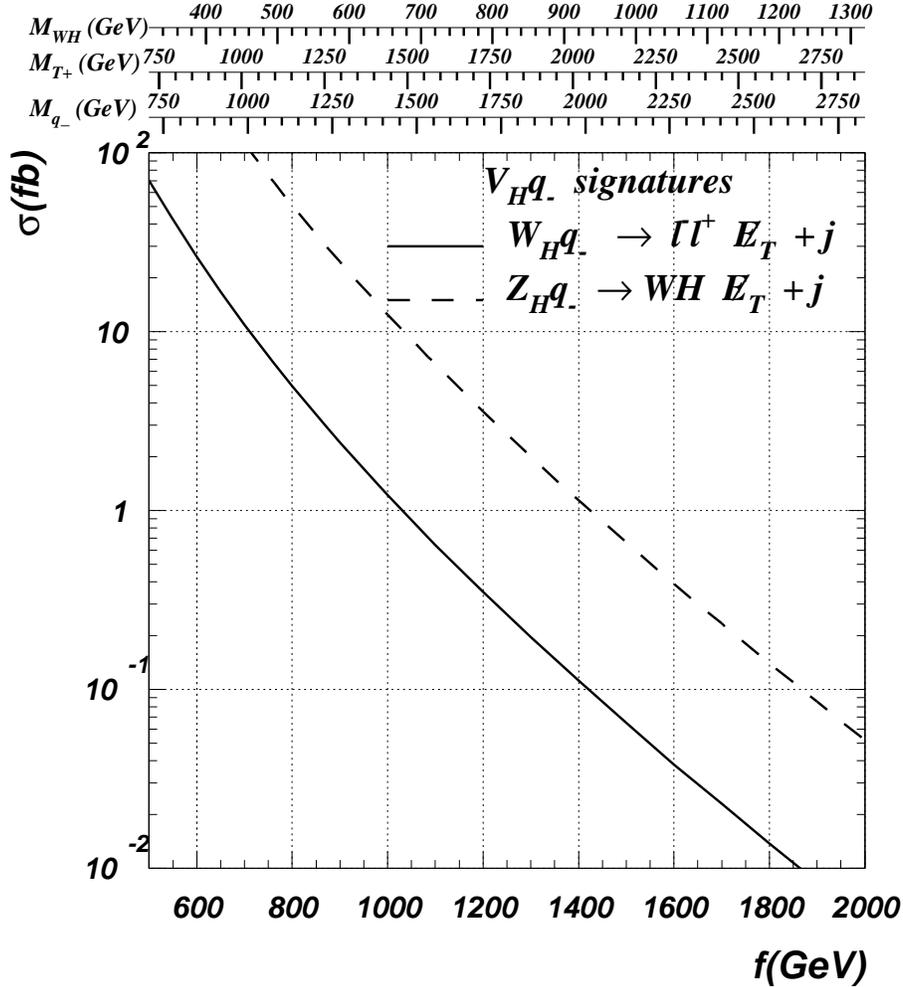}}%
\caption{\label{vqsign}
            Rates for opposite-sign lepton and 
	    associate Higgs production signatures 
            from T-odd boson and T-odd quark associate production,  $V_H q_{-}$,
            at the LHC.}
\end{figure}

\noindent
{\bf 1)  $q_-W_H$ production} \\
This signal process gives rise to OSL and 1L signatures with
one less jet as compared to the T-odd heavy quark pair production,
but without the LSL signature. The OSL signature rate for this process is presented 
as the  solid line in Fig.~\ref{vqsign}.
\\
{\bf 2) $q_-Z_H$ production} \\
The interesting decay chain of this signal process is $q_-Z_H\to
q' W_H  Z_H \to q' W^+ A_H  A_H H$ in which a high $P_T$ Higgs
boson is associatively produced with a $W$-boson. Its event rate
is large, at about 12\,fb level for $f=1$\,TeV and $\kappa=1$. With
a large $\etmiss$ in the event, it could be detectable, though a
detailed Monte Carlo study is needed.
The respective rate for  $q' W^+ A_H  A_H H$ signature is presented in
Fig.~{~\ref{vqsign} by the dashed line.
\\
{\bf 3) $q_-A_H$ production} \\
The decay chain $q_-A_H\to W_H q' A_H \to W A_H q' A_H  $ provides
the  $W^\pm$ + $\etmiss$ signature which is however not a
promising channel to look for the signal, because  the  SM
backgrounds, such as the $WZ(\to\nu\nu)$ production, could be quite  large.

\subsubsection{ $V_H V_H$  production}

The experimental signatures of $V_H V_H$ events are similar to
that of $ q_{-}V_H$ events, but with one less high-$P_T$ jet.
Therefore, it requires a larger production cross section to detect
such a signal event.
\begin{figure}[htb]
\centering{
\includegraphics[width=0.8\textwidth]{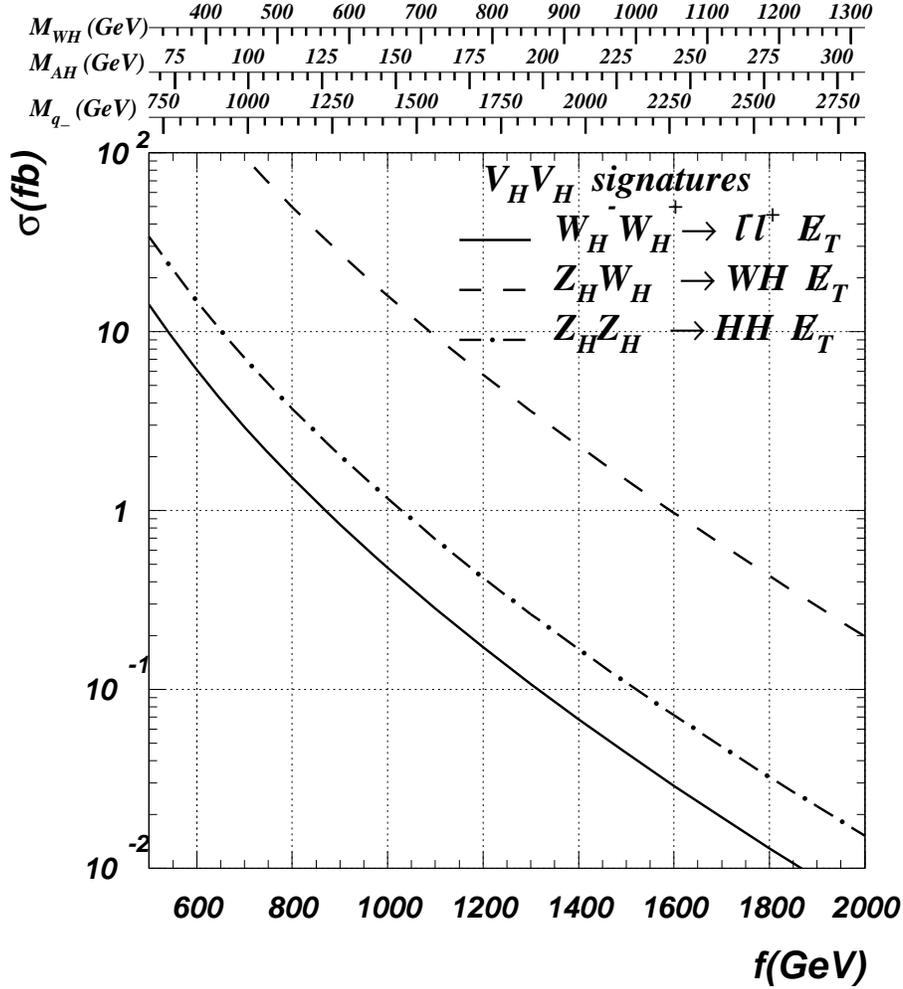}}
\caption{\label{vvsign}
            Rates for  OSL,  $WH$ and $HH$ signatures for  
            various $V_H V_H$  production reactions at the LHC.}
\end{figure}
 
\noindent
{\bf 1) $Z_H W_H$ production} \\
The event rate of  $Z_H W_H\to  A_H H W A_H$ is about the same as
that for  $ q_{-}V_H$ production, but with almost 100\% decay
branching ratio. The rate as  a function of 
$f$ is presented in  Fig.~\ref{vvsign}
by the dashed line.
Its experimental signature is the $WH$ associate
production with large missing $\etmiss$.
\\
{\bf 2) $W_H^+ W_H^-$ production} \\
The event rate of $W_H^+ W_H^-\to  W^+ A_H W^-  A_H$ is about 5
times smaller than that for $ q_{-}V_H$ production.
The solid line of  Fig.~\ref{vvsign} presents this OSL signature
rate. Hence, it is
more challenging to detect such a signal event  in the pure leptonic channel.
\\
{\bf 3) $Z_H Z_H $ production} \\
The event signature of $Z_H Z_H \to   A_H H A_H H $ is the
production of a pair of Higgs bosons with large $\etmiss$ in the
event. Its production rate is about one order of magnitude smaller
than the $W_H^+ W_H^-$ production rate,
as indicated by the dot-dashed line in Fig.~\ref{vvsign}.
On the other hand, in spite of its
small production rate, this process offers an interesting
production channel for Higgs boson pairs.

\subsubsection{Heavy T-odd Higgs boson production}

The highest heavy T-odd Higgs 
production rate, with its cross section around 1\,fb for $f\simeq$1~TeV,
comes from the $\phi^{++}\phi^{-}$ or $\phi^{--}\phi^{+}$ production channels.
For the model parameters under study, there is no allowed two-body
decay mode for  $\phi^{++}$ boson, due to mass
constraints. Nevertheless, 3-body decay modes of $\phi^{++}$ can take
place at tree level, and it is also possible to have 2-body
radiative decay modes dominating the decay branching ratios of $\phi^{++}$. 
Hence, there will be multiple jets and
leptons in such kind of signal events.

\newpage
\section{Conclusions}

The Littlest Higgs model with T-parity
(LHT)~\cite{Low:2004xc,Hubisz:2004ft,Hubisz:2005tx,Cheng:2003ju}
is an attractive Little Higgs model which provides not only
solution to the Little hierarchy problem but also a possible dark
matter candidate \cite{Asano:2006nr}. Because of the T-parity,
T-odd gauge bosons contribute to the electroweak observables at
the weak scale in pairs, hence, the mass scale ($f$) of new
particles predicted in this model can be as low as
500\,GeV~\cite{Hubisz:2005tx}. With the possibility of such a low
mass scale, many interesting phenomenology has been studied in the
literature~\cite{Hubisz:2004ft}. In order to implement T-parity in
the fermion sector of the model, the heavy T-odd $SU(2)$-doublet
fermions, which are T-parity partners of the SM fermion doublets,
have to be introduced. A preliminary study on the phenomenology of
these T-odd $SU(2)$-doublet fermions in the LHT was reported in
Ref.~\cite{sasha_pheno}. In this paper, we present a detailed
study of the phenomenology of the LHT with the emphasis on the
role of the T-odd fermions in high energy scattering processes at
the LHC.

In Sec. II, we present the effective Lagrangian of the LHT studied in
this paper.
Some relevant Feynman rules are also summarized in  Appendix A.
 Particular attention has been given to discussing the
properties of those T-odd $SU(2)$-doublet fermions (denoted as
$q_-$), including their masses and interactions to other particles
predicted in the model. As shown in Eq.~(\ref{Todd_mass}), their masses are
about ${\sqrt 2} \kappa f$ where $\kappa$ is the coefficient
introduced in the interaction Lagrangian, cf. Eq.~(\ref{kappa}), for
generating a large mass to $q_-$. (For simplicity, we have assumed
a constant $\kappa$ value, independent of quark flavor and
family.) While the typical cutoff scale of the effective theory is
about $4 \pi f$, the mass of the T-odd $SU(2)$-doublet fermions is
bounded from above due to the low energy constraint induced by the
four-fermion contact interaction presented in   Eq.~(\ref{T-odd-limit}). 
Hence, in our study
the $\kappa$ value is required to be less than about 3.4 for $f$
about 1\,TeV. Similar kind of constraints on the two parameters
($\lambda_1$ and $\lambda_2$) in the top quark sector are also
discussed in Sec. II. The results are given in Eqs.~(\ref{lambda1}), 
(\ref{unitarity}) and (\ref{naturalness}). A detailed discussion on
 the $J=1$ partial-wave amplitudes in the coupled system
of $(t\bar{t},~T_+\bar{T}_+,~b\bar{b},~WW,~Zh)$ states, which are
relevant to the top Yukawa coupling, is presented in Appendix B.
Finally, we note that the T-parity symmetry is correctly
implemented in our effective Lagrangian such that the 
$[SU(2)\times U(1)]^2$ gauge symmetry is non-linearly realized in all
sectors, including the bottom quark Yukawa interaction sector.
(It differs from the model presented in
Ref.~\cite{Hubisz:2004ft} in which the similar sector is not gauge
invariant.)

In Sec. III, we stress the importance of the T-odd $SU(2)$ doublet
fermion contributions to high energy processes. As an example, we
discuss the high energy behavior of $u\bar{u} \rightarrow W_H^+
W_H^-$ through partial-wave analysis. To have a gauge-invariant
amplitude, both types of Feynman diagrams shown in
Fig.~\ref{uubar-WHWH} need to be included in order to satisfy unitarity condition
in high energy collision. Furthermore, as shown in Sec. II, the
mass of $q_-$ is bounded from above by low energy data in this effective theory.
Therefore, the decoupling limit of the T-odd fermions is not a
realistic assumption and the T-odd fermion contribution generates
important correction to $u\bar{u}\rightarrow
W_{H}^{+}W_{H}^{-}$ process, which has not been taken into account
in the previous study~\cite{Hubisz:2004ft}. 

Because the mass of the T-odd $SU(2)$ doublet fermions cannot be
too heavy, cf. Eq.~(\ref{T-odd-limit}), they can be copiously
produced at the LHC. Therefore, in Sec. IV, we study the collider
phenomenology of the LHT with emphasis on the contributions of the
T-odd fermion to the production of the heavy T-parity partners
(either bosons or fermions) at the LHC. Fig.~\ref{qq} shows the
production cross sections of the first and second generation heavy
T-odd quarks. As shown, their cross sections are quite sizable at
the LHC. In Fig.~\ref{q3q3}, we show the similar plot for the third
generation heavy T-odd and T-even quarks. As discussed in Sec. II, the LHT
contains additional T-odd and T-even heavy quarks which are the
T-parity partners of top quark. The T-even partner ($T_+$) of top
quark can be produced singly or in pairs. 
We
also show in Fig.~\ref{vp} the associate production cross section of
T-odd fermions and T-odd gauge bosons. The associate production
cross sections of $T_+$ and weak gauge bosons are also shown  in
the same figure. In Fig.~\ref{vv}, we show the production cross sections
of heavy T-odd gauge boson pairs. As an illustration, the
dependence of the $pp\to W_H^{+} W_H^{-}$ production on the mass
of T-odd fermion is given in Fig.~\ref{whwhkappa}. For completeness, we also
show the production rate of heavy T-odd Higgs bosons in Fig.~\ref{phiphi}, though
their production rates are generally small.

Before we discuss the probable experimental signatures predicted
by this model at the LHC, we presented typical decay branching
channels of the T-parity partners in Table~\ref{table} and Fig.~\ref{tpdecay}. It
turns out that the result of Table~\ref{table}  is not very sensitive to the
choice of model parameters such as $f$, $\lambda_1$, $s_\alpha$ and $M_H$
for T-odd fermions.
Similarly, Fig.~\ref{tpdecay} also presents 
a typical pattern of branching ratios
for T-even heavy top-quark partner $T_+$,
though this pattern depends on the choice of model parameters  
such as $s_\alpha$ and $M_H$.
 
Combining the
information on the production cross section of heavy particles and
their decay branching ratios into particular decay channels, one
can easily calculate   production rates of events with certain
experimental signature. Some of those results are shown in the
remaining figures of the paper.

We concluded in Sec.~IV that the like-sign di-lepton signature of
the 1st and 2nd generation heavy T-odd quark pair production is the
most useful channel to discover these new have quarks at the LHC.
Because the heavy T-odd gauge boson $Z_H$  almost  always decays
into a pair of Higgs boson $H$ and T-odd photon $A_H$, the
production processes with $Z_H$ in the final state provide  a new
production mechanism for single-Higgs or Higgs-pair production.
Their rates are presented in Figs.~\ref{vqsign}~and~\ref{vvsign}, respectively.

We also provide for the first time
the complete  CalcHEP LHT model files 
including T-odd  $SU(2)$ doublet fermions,
which is available at
\verb|http://hep.pa.msu.edu/LHT/|.

\acknowledgments
We thank Qing-Hong Cao, Jay Hubisz,
Alexander Pukhov and Riccardo Rattazzi for useful
discussions. C.P.Y. and K.T. thank the National Center for
Theoretical Sciences in Taiwan for its hospitality, where part of
the work was done. This work was supported in part by the US
National Science Foundation under award PHY-0555545.
\vspace{1in}

{\bf Note added:} While finalizing the write-up of this work,
we are aware of the paper by A.~Freitas and
D.~Wyler~\cite{Freitas:2006vy} who studied the phenomenology
of T-odd fermion in the LHT.

\appendix
\section{Feynman rules}
Most of the Feynman rules  for the Littlest Higgs model with T-parity have
been presented in Ref.~\cite{Hubisz:2004ft} and the references
therein except for T-odd $SU(2)$ doublet fermions. We agree with
their results in the hep-ph arXiv version (v3) of
Ref.~\cite{Hubisz:2004ft} after identifying  $T_+=-t'_+$ and $T_-=-t'_-$
(their $t'_\pm$ fields correspond to our $T_\pm$). In this
appendix, we list a few Feynman rules relevant to our analysis,
especially those related to T-odd fermions.
They are listed in Tables~\ref{feynman-rules1}--\ref{feynman-rules3} below.

We also provide for the first time the complete  CalcHEP LHT
model files  including T-odd  $SU(2)$ doublet fermions which
is available at
\verb|http://hep.pa.msu.edu/LHT/|.

In Tables IV, V and VI, we have defined the following coefficients.
  $s_H~(=\sin\theta_H)$ describes the degree
  of mixing between heavy neutral gauge bosons with
$s_H\simeq \frac{gg'}{g^2-g^{'2}/5}\frac{v_{SM}^2}{4f^2}$
and $c_H=\cos\theta_H$.
Also,
$s_L \simeq s_\alpha^2 \frac{v_{SM}}{f}$ and
$c_L =\sqrt{1-s_L^2}$. In addition,
$P_{L}=\frac{1-\gamma_{5}}{2}$ and $P_{R}\
=\frac{1+\gamma_{5}}{2}$
are the left-handed and right-handed projection operators, respectively.
We note that in those tables we have suppressed the CKM matrix element
dependence. For example, from Table VI, we can read out the coupling of
$W^+_\mu \bar{t} b$ to be $V_{tb} (i\frac{g}{\sqrt{2}}c_L \gamma_\mu
P_L)$, after restoring the CKM matrix element $V_{tb}$ derived from the
interaction Lagrangian.
In the above expression, the product of $V_{tb} c_L$, which is defined
as $V_{tb}^{eff}$, should be identified with the CKM matrix element
determined from the low energy processes (or from measuring the SM
single-top direct production rate at the Tevatron or the LHC~\cite{Yuan:1989tc,single_top_t}).
Thus, from Table VI, we read out the coupling of
$W^+ \bar{T}_+ b$
to be
$V_{tb} (i\frac{g}{\sqrt{2}}s_L\gamma_\mu P_L)$,
after restoring the CKM matrix element dependence, which can be
rewritten as
$V_{tb}^{eff} (i\frac{g}{\sqrt{2}}{\frac{s_L}{c_L}}\gamma_\mu P_L)$.
The coefficient of $W^+\bar{T}_+ b$ coupling $V_{tb}^{eff}{\frac{s_L}{c_L}}$
is approximately equal to $V_{tb}^{eff}s_L$ up to $v_{SM}^2/f^2$ corrections,
for $s_L\propto v_{SM}/f$, cf. Eq.~(\ref{s_LR}).

\begin{table}[h]
\begin{center}
\begin{tabular}{|c|c||c|c|}
\hline
Interaction & Feynman rule & Interaction & Feynman rule\\
\hline \hline
$W_{H_\mu}^+ \bar{u} d_-$ & $i\frac{g}{\sqrt{2}}\gamma_\mu P_L$ & $W_{H_\mu}^- \bar{d} u_-$ & $i\frac{g}{\sqrt{2}}\gamma_\mu P_L$\\
$Z_{H_\mu} \bar{u} u_-$ & $i(\frac{g c_H}{2}-\frac{g' s_H}{10}) \gamma_\mu P_L$ &
$Z_{H_\mu} \bar{d} d_-$ & $i(-\frac{g c_H}{2}-\frac{g' s_H}{10}) \gamma_\mu P_L$ \\
$A_{H_\mu} \bar{u} u_-$ & $i(-\frac{g s_H}{2}-\frac{g' c_H}{10}) \gamma_\mu P_L$ &
$A_{H_\mu} \bar{d} d_-$ & $i(\frac{g s_H}{2}-\frac{g' c_H}{10}) \gamma_\mu P_L$ \\
\hline 
\end{tabular}
\caption{Feynman rules for the 1st and 2nd 
generation T-odd fermion interaction with heavy gauge boson and
SM fermion.\label{feynman-rules1}}
\end{center}
\end{table}

\begin{table}[h]
\begin{center}
\begin{tabular}{|c|c||c|c|}
\hline
Interaction & Feynman rule & Interaction & Feynman rule\\
\hline \hline
$W_{H_\mu}^+ \bar{t} b_-$ & $i\frac{gc_L}{\sqrt{2}}\gamma_\mu P_L$ & $W_{H_\mu}^- \bar{b} t_-$ & $i\frac{g}{\sqrt{2}}\gamma_\mu P_L$\\
$W_{H_\mu}^+ \bar{T}_+ b_-$ & $i\frac{g s_L}{\sqrt{2}}\gamma_\mu P_L$ & & \\
$Z_{H_\mu} \bar{t} t_-$ & $i(\frac{g c_H}{2}-\frac{g' s_H}{10})c_L \gamma_\mu P_L$ &
$Z_{H_\mu} \bar{b} b_-$ & $i(-\frac{g c_H}{2}-\frac{g' s_H}{10}) \gamma_\mu P_L$ \\
$Z_{H_\mu} \bar{T}_+ t_-$ & $i(\frac{g c_H}{2}-\frac{g' s_H}{10})s_L \gamma_\mu P_L$ &
$Z_{H_\mu} \bar{t} T_-$ & $-i\frac{2}{5}g' s_H \gamma_\mu (s_L P_L +s_R P_R)$ \\
$Z_{H_\mu} \bar{T}_+ T_-$ &  $i\frac{2}{5}g's_H \gamma_\mu (c_L P_L +c_R P_R)$& & \\
$A_{H_\mu} \bar{t} t_-$ & $i(-\frac{g s_H}{2}-\frac{g' c_H}{10})c_L \gamma_\mu P_L$ &
$A_{H_\mu} \bar{b} b_-$ & $i(\frac{g s_H}{2}-\frac{g' c_H}{10}) \gamma_\mu P_L$ \\
$A_{H_\mu} \bar{T}_+ t_-$ & $i(-\frac{g s_H}{2}-\frac{g' c_H}{10})s_L \gamma_\mu P_L$ &
$A_{H_\mu} \bar{t} T_-$ & $-i\frac{2}{5} g' c_H\gamma_\mu (s_L P_L+s_R P_R)$ \\
$A_{H_\mu} \bar{T}_+ T_-$ & $i\frac{2}{5} g' c_H\gamma_\mu (c_L P_L+c_R P_R)$ & &\\
\hline 
\end{tabular}
\caption{Feynman rules for the 3rd generation
 T-odd fermion interaction with heavy gauge boson and
SM fermion.\label{feynman-rules2}}
\end{center}
\end{table}

\begin{table}[h]
\begin{center}
\begin{tabular}{|c|c||c|c|}
\hline
Interaction & Feynman rule & Interaction & Feynman rule\\
\hline \hline
$W^+_\mu \bar{t} b$ & $i\frac{g}{\sqrt{2}}c_L \gamma_\mu P_L$ & $W^+ \bar{T}_+ b$ & $i\frac{g}{\sqrt{2}}s_L\gamma_\mu P_L$\\
$Z_\mu \bar{t} t$ & $i\frac{g}{c_W}\gamma_\mu \left\{(\frac{1}{2}c_L^2-\frac{2}{3}s_W^2) P_L -\frac{2}{3}s_W^2 P_R\right\}$ 
& $Z_\mu \bar{t} T_+$ & $i\frac{g}{c_W}\frac{s_L c_L}{2} \gamma_\mu P_L$\\
$Z_\mu \bar{T}_+ T_+$ & $i\frac{g}{c_W}\gamma_\mu \left\{(\frac{1}{2}s_L^2-\frac{2}{3}s_W^2) P_L -\frac{2}{3}s_W^2 P_R\right\}$ &  & \\
\hline 
\end{tabular}
\caption{Feynman rules for the SM gauge interaction with top sector.\label{feynman-rules3}}
\end{center}
\end{table}

\section{Unitarity bound from $J=1$ partial wave amplitudes in the coupled system of 
($t\bar{t}$, $T_+\bar{T}_+$, $b\bar{b}$, $WW$, $Zh$) states}

The amplitudes for $t\bar{t}\rightarrow t\bar{t}$, $T_+\bar{T}_+$, $b\bar{b}$, $WW$, and $Zh$ processes
and their inverse processes
contribute to $J=1$ partial wave amplitude matrix in this coupled system. 
The $J=1$ partial wave amplitudes are given by
\begin{eqnarray}
a_{\mu\mu'}^1 &=& \frac{1}{32\pi}\int_{-1}^1 d(\cos\theta)d^1_{\mu\mu'}(\theta)T_{\mu\mu'}.
\end{eqnarray}
Here $d_{\mu\mu'}^1(\theta)$ is the well-known Wigner d-function. For fermions, $\mu$ and 
$\mu'$ are defined by $\mu=(\lambda-\bar{\lambda})/2$ and $\mu'=(\lambda'-\bar{\lambda}')/2$, 
where $\lambda$'s are the helicities of the fermions: $\lambda~(\bar{\lambda})$ for the initial 
state fermion (anti-fermion) and $\lambda'~(\bar{\lambda}')$ for the final state fermion 
(anti-fermion), and for bosons, $\mu=0$. $T_{\mu\mu'}$ is a helicity amplitude with $\mu$ and $\mu'$.

Writing the channels in the order $t_+\bar{t}_-$, $(T_+)_+(\bar{T_+})_-$, $W^+W^-$, $hZ$, 
$t_-\bar{t}_+$ and $b_-\bar{b}_+$, the $J=1$ partial wave amplitude matrix $a^1$ is given by
\begin{equation}
a^1 = \frac{M_t^2}{16\pi v_{SM}} \left(
\begin{array}{cccccc}
0 & 0 & -\sqrt{2}     & -i\sqrt{2}     & - 1 &  +1\\
0 & 0 & -\sqrt{2}R^2  & -i\sqrt{2}R^2  & R^2 & R^2\\
-\sqrt{2} & -\sqrt{2}R^2 & 0 & 0 & 0 & \sqrt{2}(1+R^2)\\
i\sqrt{2} & i\sqrt{2}R^2 & 0 & 0 & i\sqrt{2}(1+R^2) & 0\\
-1 & R^2 & 0 & -i\sqrt{2}(1+R^2) & 0 & 0\\
1 & R^2 & \sqrt{2}(1+R^2) & 0 & 0 & 0\\
\end{array}
\right),
\label{J1_amp}
\end{equation}
where $R=\lambda_1/\lambda_2$. Here we have assumed that the center-of-mass energy $\sqrt{s}$ is much larger
than masses of particles considered here, and only couplings in top sector are relevant, and gauge couplings 
and all other Yukawa couplings are taken to be zero. 
We have not shown explicitly the color indices in Eq.~(\ref{J1_amp}), however all color neutral channels should 
be taken into account. Thus the $J=1$ partial wave amplitude matrix in this system is $14\times 14$.
Note that the parameter $R$ is the only unknown parameter in Eq.~(\ref{J1_amp}), 
and the absolute value of the largest eigenvalue of the $J=1$ partial wave amplitude matrix 
increases as $R$ gets larger. The requirement that the absolute value of the largest eigenvalue 
be less than a half $(|a^1_{\rm max}|<1/2)$ yields the upper bound on the parameter R as
\begin{eqnarray}
R<3.3,
\end{eqnarray}
for $M_t=175$ GeV.
In terms of $s_\alpha$, this bound corresponds to
\begin{eqnarray}
s_\alpha<0.96,
\end{eqnarray}
since $R=s_\alpha/c_\alpha$. Using the top-quark mass constraint, cf. Eq.~(\ref{lambda1}), this bound
generates a bound on $\lambda_1$ as
\begin{eqnarray}
\lambda_1=\frac{M_t}{v_{SM}\sqrt{1-s_\alpha^2}}<2.5,
\end{eqnarray}
for $M_t=175$ GeV.

\end{document}